\documentclass[a4paper,twocolumn,superscriptaddress,prx, aps, floatfix ]{revtex4-2}
\usepackage{latexsym,amsmath}
\usepackage{amsbsy}
\usepackage[pdftex]{graphicx}
\usepackage{amssymb}
\usepackage{epstopdf}
\usepackage{verbatim}

\usepackage[usenames]{color}
\usepackage{capt-of} 
\usepackage{float}
\usepackage{ esint }
\usepackage{hyperref}
\usepackage{graphicx}
\usepackage[normalem]{ulem}

\usepackage{xcolor}

\begin{document}

\title{Random graphs and real networks with weak geometric coupling}

\author{Jasper van der Kolk}
\email{jasper.vanderkolk@ub.edu}
\affiliation{Departament de F\'isica de la Mat\`eria Condensada, Universitat de Barcelona, Mart\'i i Franqu\`es 1, E-08028 Barcelona, Spain}
\affiliation{Universitat de Barcelona Institute of Complex Systems (UBICS), Barcelona, Spain}

\author{\\M. \'Angeles \surname{Serrano}}
\email{marian.serrano@ub.edu}
\affiliation{Departament de F\'isica de la Mat\`eria Condensada, Universitat de Barcelona, Mart\'i i Franqu\`es 1, E-08028 Barcelona, Spain}
\affiliation{Universitat de Barcelona Institute of Complex Systems (UBICS), Barcelona, Spain}
\affiliation{Instituci\'o Catalana de Recerca i Estudis Avan\c{c}ats (ICREA), Passeig Llu\'is Companys 23, E-08010 Barcelona, Spain}

\author{Mari\'an Bogu\~n\'a}
\email{marian.boguna@ub.edu}
\affiliation{Departament de F\'isica de la Mat\`eria Condensada, Universitat de Barcelona, Mart\'i i Franqu\`es 1, E-08028 Barcelona, Spain}
\affiliation{Universitat de Barcelona Institute of Complex Systems (UBICS), Barcelona, Spain}

\begin{abstract}
Geometry can be used to explain many properties commonly observed in real networks. It is therefore often assumed that real networks, especially those with high average local clustering, live in an underlying hidden geometric space. However, it has been shown that finite size effects can also induce substantial clustering, even when the coupling to this space is weak or non existent. In this paper, we study the weakly geometric regime, where clustering is absent in the thermodynamic limit but present in finite systems. Extending Mercator, a network embedding tool based on the Popularity$\times$Similarity $\mathbb{S}^1$/$\mathbb{H}^2$ static geometric network model, we show that, even when the coupling to the geometric space is weak, geometric information can be recovered from the connectivity alone. The fact that several real networks are best described in this quasi-geometric regime suggests that the transition between non-geometric and geometric networks is not a sharp one.
\end{abstract}

\maketitle

\section{Introduction}
Over the past few decades, the use of complex networks to describe the properties of systems of many interacting particles has become widespread in many fields of science. Examples of such fields include ecology~\cite{Dunne2005}, neurology~\cite{Sporns2013}, the social sciences~\cite{Freeman2004} and technology~\cite{Barabasi2001}. Real networks from all these fields have been found to share several important properties, such as the small-world property~\cite{Buchanan2003}, scale-free degree distributions~\cite{Albert2002}, and high levels of clustering~\cite{Newman2003}. The search for models which can reproduce these features has led to the field of network geometry~\cite{Boguna2021}, and more specifically to the introduction of geometric random graphs. In this class of models, nodes are assumed to live in an underlying metric space that conditions the connectivity of the graph. The fact that this approach is able to reproduce all the basic network properties and symmetries~\cite{Serrano2008,Muller2019,Friedrich2018,Fountoulakis2021,Candellero2016,Gugelmann2012,Krioukov2010} as well as produce strong results in practical tasks such as community detection~\cite{Serrano2012,GarciaPerez2016}, information routing~\cite{Allard2020,Gulyas2015,Boguna2009} and link prediction~\cite{Wang2016,Kitsak2020} has led many to wonder if there is a way to determine if real networks are indeed geometric in nature~\cite{Krioukov2016,Michielan2022}. In general, (some function of) the amount of closed triangles in the network, expressed, for example, by the average local clustering coefficient, is taken to indicate the presence of geometry. However, these studies do not contemplate the fact that the transition between non-geometric and geometric networks might not be sharp. Indeed, in Ref.~\cite{vanderKolk2022} it was shown that there exists a quasi-geometric region where the coupling of a network to its underlying metric space is weak. Here, even though in the thermodynamic limit the average local clustering coefficient vanishes, its decay with the system size is extremely slow, which means that for finite systems the level of clustering remains significant. \\
This observation raises several questions, the most important being what it implies for real networks that might live in this regime. How strongly does the underlying metric space affect the topology of networks when the coupling is weak? \\
We study how well the original coordinates of the nodes, used to generate a network, can be reproduced from the topology alone. Being able to reproduce the coordinates with high precision means that information about the geometry is contained in the topology, and one can thus say that the network is truly geometric. Conversely, if the coordinates cannot be found, it is a good indication that the connection between the geometry and the topology is too weak for the network to be considered geometric. \\
To study this we extend Mercator~\cite{GarciaPerez2019}, a network embedding tool that uses a combination of machine learning as well as maximum likelihood techniques to recover the hidden coordinates of the nodes in a network from its connectivity. It is based on the $\mathbb{S}^1$-model, the only maximal entropy random geometric network model that can reproduce the small-world property, heterogeneous degree distributions, and high levels of clustering. However, as is the case with many network embedding tools~\cite{Papadopoulos2015,Blasius2018,Muscoloni2017}, Mercator does not contemplate the possibility of quasi-geometric networks, i.e., those networks where the coupling to the underlying metric space is so weak that clustering vanishes in the thermodynamic limit. In this paper, we extend Mercator to this regime. Embedding these type of networks, we find that for a range of weak couplings the original geometry can indeed be recovered based solely on the topology. We also show that this region coincides with the quasi-geometric regime defined in Ref.~\cite{vanderKolk2022} on the basis of the scaling properties of the average local clustering coefficient in the $\mathbb{S}^1$-model. Additionally, it is found that Mercator can provide meaningful embeddings even when the network is explicitly non-geometric, even in the case of the configuration model~\cite{vanderHoorn2018}. Similarly to how fluctuations can induce spurious communities~\cite{Guimera2004,Peixoto2022}, finite size effects can lead to an \textit{effective geometry}, which can be used, for example, in greedy routing algorithms.\\
Finally, we show that some real networks are best described in the quasi-geometric regime, where the embedding is capable of reproducing the topological properties of the network accurately.

\section{Methods}
\subsection{The $\mathbb{S}^1$/$\mathbb{H}^2$-models}
We base this paper on the $\mathbb{S}^1$ geometric network model~\cite{Serrano2008}, as well as on its isomorphic equivalent, the $\mathbb{H}^2$ model~\cite{Krioukov2010}. In the $\mathbb{S}^1$-model, nodes are assumed to live in a metric similarity space, where similarity encodes for all attributes of a node, apart from its degree. This similarity space is given by a circle of radius $R$, where each node $n$ has an angular coordinate $\theta_n\sim \mathcal{U}(0,2\pi)$. Note that the angular distribution does not necessarily need to be uniform, and choosing another distribution leads to the creation of soft communities~\cite{GarciaPerez2018b}. The degree of a node is determined by its popularity, which in the $\mathbb{S}^1$-model is represented by an internal parameter $\kappa_n$ drawn from some distribution $\rho(\kappa)$. We then impose that the probability of two nodes being connected resembles a gravity law. Nodes that are further away in similarity space are less likely to be connected, whereas nodes with higher popularity are more likely to be connected. Finally, we want the model to define an ensemble of geometric random graphs that maximizes entropy. This fixes the connection probability up to a single free parameter that sets the level of randomness in the system and can thus be thought of as a temperature~\cite{Krioukov2009}. The functional form of the connection probability is given by 
\begin{equation}
	p_{nm} = 
	\left(1+\dfrac{x_{nm}^\beta}{\left(\hat{\mu}\kappa_n\kappa_m\right)^{\max(1,\beta)}}\right)^{-1}\label{eq:pij_S1},
\end{equation}
where $x_{nm}$ is the distance in similarity space between the two nodes $n$ and $m$. In the $\mathbb{S}^1$-model, this is the distance along the circle given by $x_{nm} = R\,(\pi-|\pi-|\theta_n-\theta_m||)$. The radius $R$ is given by $R=N/(2\pi)$, such that in the thermodynamic limit, $N\rightarrow\infty$, the distribution of nodes is given by a Poisson point process along the real line with rate one. We set the parameter $\hat\mu$ such that the expected degree of a node with hidden degree $\kappa$ is given by $\overline k(\kappa) = \kappa$. This allows us to interpret the hidden degree as the expected degree of a node. In the thermodynamic limit one can give a closed form for $\hat\mu$. For $\beta>1$ one has $\hat\mu = \beta \sin\left(\pi/\beta\right)/(2\pi\langle k\rangle)$ whereas for $\beta<1$ the parameter becomes size dependent: $\hat\mu = (1-\beta)N^{\beta-1}/(2^\beta\langle k\rangle)$. When working close to $\beta=1$, second order terms become important and the approximations above break down for finite systems and one needs to find $\hat\mu$ numerically. \\
The parameter $\beta = 1/T$ is the inverse temperature of the system. The temperature $T$ represents the coupling strength between the geometry and topology, i.e. how strictly the connectivity of points is dictated by their coordinates in similarity space. At zero temperature, the model is completely deterministic, which means that a certain distribution of $\theta$'s and $\kappa$'s leads to exactly one network realization. At infinite temperature, the role of the underlying metric space is completely lost. Note that we want the connection probability to remain dependent on the hidden degrees, even when $\beta\rightarrow0$. This way, the model reduces to the hyper-soft configuration model~\cite{vanderHoorn2018} in the infinite temperature limit. The coupling strength between the similarity space and the topology can be associated with the level of clustering in the system, and it has been shown that at the critical beta $\beta_c=1$ the system transitions from a region of high clustering at high $\beta$ to a region of vanishing clustering at low $\beta$~\cite{Serrano2008}. This can be understood as follows: As the underlying similarity space is metric, the triangle inequality holds; if a node lies close to two of its neighbors, these must necessarily also lie close together. When the coupling is strong, this effect translates over to the topology of the network, thus leading to many triangles. When the coupling is weak, connection become long range and the effect of triangle inequality diminishes, leading to less triangles. \\
The $\mathbb{H}^2$-model is the hyperbolic equivalent of the $\mathbb{S}^1$-model. Here, the underlying metric space is not an 1-dimensional sphere, but rather the 2-dimensional hyperbolic plane, where the radial coordinate encodes the popularity dimension. As there exists an isomorphism between the hidden degree $\kappa$ and the radial degree $r$, the two models are isomorphic to one another. \\
In the hyperbolic plane, distances between points are defined by the hyperbolic law of cosines:
\begin{alignat}{2}
	\cosh \big(\zeta x_{nm}\big)= &\cosh \big(\zeta r_n\big) \cosh \big( \zeta r_m\big) \nonumber\\
	- &\sinh \big( \zeta r_n\big) \sinh \big(\zeta r_m\big) \cos \Delta\theta_{nm},
\end{alignat}
where $\zeta = \sqrt{-K}$ and $K$ is the constant, negative curvature of the hyperbolic disk.  It has, however, been shown that the distance can be approximated as 
\begin{equation}
	x_{nm} \approx r_n + r_m + \frac{2}{\zeta} \ln \frac{\Delta\theta_{nm}}{2}\label{eq:hypdist_approx},
\end{equation}
where the amount of node pairs for which this approximation fails vanishes in the thermodynamic limit~\cite{Serrano2022}. The connection probability that maximizes entropy while fixing the expected degree distribution is given by 
\begin{equation}
	p_{ij} = \left(1+e^{\frac{\beta\zeta}{2}(x_{nm}-R_{\mathbb{H}^2})}\right)^{-1}\label{eq:pij_H2},
\end{equation}
where $R_{\mathbb{H}^2}$ is the radius of the hyperbolic disk and $\zeta$ is the curvature.
In order to find the transformation between both models, we must equate the connection probabilities given in Eqs.~\eqref{eq:pij_S1} and \eqref{eq:pij_H2}. This then leads to the following transformation
\begin{equation}
	\kappa(r) = \kappa_0\,\exp\left(\frac{\beta\zeta}{2\max(1,\beta)}\big(R_{\mathbb{H}^2}-r\big)\right),
\end{equation}
where we have defined the radius such that $\kappa(R_{\mathbb{H}^2}) = \kappa_0$, the smallest possible hidden degree. Nodes with larger degrees thus necessarily lie closer to the center of the disk. In this case, the hyperbolic radius is given by 
\begin{equation}
	R_{\mathbb{H}^2} = \frac{2}{\zeta}\ln\left(\frac{N}{\pi}\right) - \frac{2\max(1,\beta)}{\beta\zeta}\ln\left(\hat{\mu}\kappa_0^2\right).
\end{equation}
Let us now take a closer look at the curvature of the hyperbolic disk. Interestingly, Eq.~\ref{eq:pij_H2} is reminiscent of the Fermi-Dirac distribution. We can think of the distance $x_{nm}$ as defining the energy of a certain link, and $R_{\mathbb{H}^2}$ acting as a chemical potential. In this picture, $2/\zeta$ is equivalent to the Boltzmann constant. Similarly to how setting this constant to one in statistical physics implies a change of units, a change in the curvature in hyperbolic space can always be reabsorbed into a change of length scale. We are thus free to choose $\zeta$ in any way that is convenient. For $\beta>1$, we choose to set $\zeta = 1$, as is typically done when studying this regime~\cite{Serrano2022}. In the case $\beta < 1$, we set $\zeta = \beta ^{-1}$, leading to an infinite negative curvature at $\beta = 0$~\cite{Krioukov2010}. \\
This choice has several advantages. First, it is the only definition that leads to a finite hyperbolic radius at $\beta = 0$, which is important as the $\mathbb{H}^2$-model is mostly used for visualization purposes. Second, it allows for an intuitive interpretation of the hyperbolic distance defined in Eq.~\ref{eq:hypdist_approx}. As $\beta\rightarrow0$, the dependence on the angular distance $\Delta\theta_{ij}$ vanishes, which is in line with the fact that this limit corresponds to the hyper-soft configuration model, where only the popularity dimension plays a role. If the curvature were set to some other value, the typical length scale in the popularity dimension would diverge whereas it would remain constant in the similarity dimension, effectively leading to the same situation. 

\subsection{Embedding of networks in the weakly geometric regime}

Network geometry has important practical implications for real systems. For example, it can be used for routing information on the Internet~\cite{Boguna2010}, for community detection~\cite{Serrano2012,GarciaPerez2016} or for the prediction of missing links~\cite{Wang2016,Kitsak2020}, as well as for creating downscaled network replicas~\cite{GarciaPerez2018a}. In order to do so, one needs to be able to faithfully embed real-world networks into the hidden metric space using only the information contained in their topology. \\
Even though there are many ways to obtain such an embedding~\cite{Papadopoulos:2015ub,Blasius:2018dx,Kitsak2020,Blasius:2020ko,Kovacs:2021zy,goyal2018graph,AlanisLobato2016,muscoloni2017machine,Keller-Ressel:2020ni}, in this paper we focus on Mercator~\cite{GarciaPerez2019}, which finds the hidden $\mathbb{S}^1/\mathbb{H}^2$ coordinates of real networks such that realizations based on these coordinates best reproduce the properties of the original network. Mercator employs a combination of machine learning and maximum likelihood methods to achieve this goal, which allows it to be both precise as well as efficient. \\
Mercator first proposes a (random) $\beta$ and sets $\hat{\mu}$ as a function of the average degree. The hidden degrees $\{\kappa_n\}$ are then set such that the expected degree $\overline{k}(\kappa_n)$ of each node coincides with the observed degree $k_n$ in the original network. This is done iteratively starting from $\kappa_n = k_n$. With this information, the expected average local clustering coefficient is calculated. If the clustering is higher (lower) than in the original network, $\beta$ is adjusted downwards (upwards). This process is then repeated until the correct level of clustering is achieved.\\
The next step uses Laplacian Eigenmaps (LE)~\cite{Belkin2008,AlanisLobato2016} to infer the initial positions of the nodes in similarity space. Mercator then makes order preserving adjustments by calculating the expected gap size in the $\mathbb{S}^1$-model. It then refines the positions by maximizing the likelihood that the observed network is generated by the model. 
The final step of Mercator refines the set of $\kappa_n$'s, now using the angular information found in the previous steps. \\
In this original version, Mercator is only able to handle strongly geometric networks with $\beta > 1$. It is a standard assumption that geometric networks must live in this regime due to their finite levels of clustering. However, as explained in the introduction, the $\mathbb{S}^1/\mathbb{H}^2$-model in the regime $0<\beta\leq1$ can produce networks with substantial clustering away from the thermodynamic limit. It is therefore necessary to extend Mercator to this new region. This implies implementing the change in connection probability when crossing the critical temperature as found in Eq.~\ref{eq:pij_S1}. The same goes for the parameter $\hat\mu$. However, here we cannot just take the other asymptotic equation for $\beta<1$ because we want Mercator to be applicable to small networks with $\beta\approx1$ as well. Hence, we produced a new version of Mercator that is able to handle networks in the whole range of $\beta$ values and where $\hat\mu$ is determined numerically such that the observed average degree of the real network matches exactly the expected average degree of a $\mathbb{S}^1$-network with uniform node distribution and a hidden degree distribution that matches the observed one. \\
\begin{figure}[t]
	\centering
	\includegraphics[width=\columnwidth]{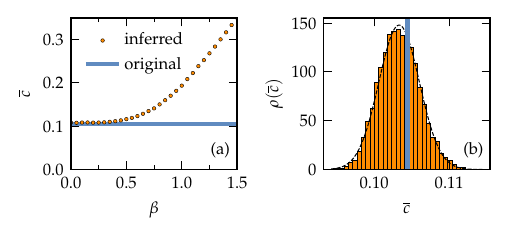}
	\vspace{-8mm}
	\caption{Inferring beta for the Ecological network (details in Tab.~\ref{tab:realnets}). \textbf{(a)} The orange points represent the inferred average local clustering coefficient given the fitted hidden degree and different betas, and the blue horizontal line indicates the original clustering coefficient of the network. \textbf{(b)} The histogram represents the probability density of the local clustering coefficient produced by degree preserving randomization of the connectivity of the original network. The black dotted line is the fitted normal distribution and the continuous blue line indicates the original clustering. }
	\label{fig:metricity}
\end{figure}
One of the challenges of embedding networks below $\beta=1$ comes from the fact that the function $\overline{c}(\beta)$ flattens off as one approaches the infinite temperature limit $\beta\rightarrow 0$, as can be seen in Fig.~\ref{fig:metricity}a. Here, we take as an example an ecological network~\cite{Dunne2014}, where nodes represent taxa and edges trophic relationships. As per the initial steps of Mercator, we choose a certain $\beta$ and set the hidden degrees such that the degree distribution is reproduced and with this calculate the expected average local clustering coefficient $\langle \overline{c}\rangle$. Repeating this for a range of $\beta$'s, one observes that the function approaches a constant as $\beta$ approaches zero. The horizontal line in the figure represents the actual level of clustering in the ecological network under study. We can say, with confidence, that $\beta_{real} \lesssim 0.5$, but cannot determine a lower bound. As the true value could be $\beta=0$, i.e. the levels of clustering in the network could be described by the configuration model, the coupling between the geometry and topology of the network is extremely weak and it is, thus, effectively non-geometric. We conjecture that these networks either have no associated geometry to begin with, or are coupled so weakly to it that it cannot be reproduced. Thus, Mercator must be able to detect these types of networks. In order to do so we want Mercator to answer the following question: ``Can the observed levels of clustering be plausibly explained by the configuration model?". \\
To answer this question, we need to add a step to the algorithm. Before the embedding of a network starts, a large amount of random copies are created using degree-preserving randomization~\cite{Cobb2003}. This randomization step destroys all information contained in the network, except for the degrees of the nodes and structural correlations imposed by global constraints at finite sizes. Because the angular coordinate in the $\mathbb{S}^1/\mathbb{H}^2$-model functions as a proxy for all attributes of a node, except for its degree, it is clear that removing this information is equivalent to decoupling the network from its similarity dimension, exactly what happens at $\beta=0$. Thus, these random copies are just realizations of the configuration model preserving the original degree distribution. We then calculate the average local clustering coefficient for all randomized copies, leading to the distribution shown in Fig.~\ref{fig:metricity}b. The observed level of clustering is given by the vertical line, and we can conclude that it is completely in line with the configuration model. Had the observed clustering been much larger, we might conclude that it is statistically unlikely that the network was generated with the configuration model and that $\beta>0$. For networks where this is the case, it is meaningful to employ Mercator to generate embeddings.

\section{Results}
\subsection{Artificial Networks. Embeddings}
\begin{figure}[t]
	\centering
	\includegraphics[width=\columnwidth]{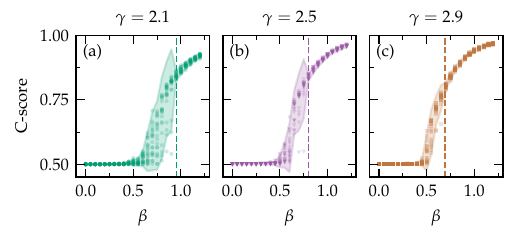}
	\caption{Dots represent the $C$-score as a function of the inverse temperature $\beta$ for individual realizations of the $\mathbb{S}^1$-model for $N=4000$, $\langle k\rangle=12$ and \textbf{(a)} $\gamma=2.1$ , \textbf{(b)} $\gamma=2.5$ and \textbf{(c)} $\gamma=2.9$. The shaded regions represent the $2\sigma$ confidence interval and the vertical dashed lines indicate the critical inverse temperature $\beta'_c=2/\gamma$ separating the quasi-geometric and non-geometric regimes. }
	\label{fig:cscore}
\end{figure}

It is important to first study if it is indeed possible to recover geometric information from the topology of a weakly geometric network when the ground truth about its geometry is known. In order to do this, we study the performance of Mercator on artificial networks generated from the $\mathbb{S}^1$-model with $0<\beta<1$. We generate heterogeneous networks where the distribution of hidden degrees is given by
\begin{equation}
	\rho(\kappa) = (\gamma-1)\kappa_0^{\gamma-1}\kappa^{-\gamma}.
\end{equation}
Here, $\gamma > 2$ determines the exponent of the tail of the resulting degree distribution.\\
We focus on Mercator's ability to recover the angular coordinates of the original network, as it is only the coupling to the similarity dimension that becomes weaker as $\beta\rightarrow0$. The performance with respect to the popularity dimension should not be much different than in the region $\beta>1$, which has been already extensively studied in Ref.~\cite{GarciaPerez2019}. \\
\begin{figure}[t]
	\centering
	\includegraphics[width=\columnwidth]{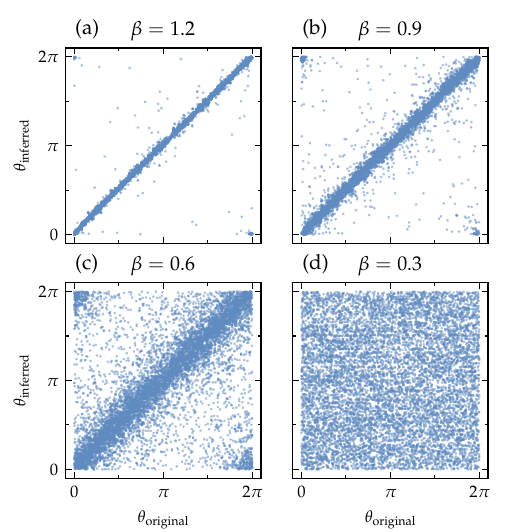}
	\caption{Examples of the inferred angular coordinate versus the original angular coordinate for artificial networks generated with the $\mathbb{S}^1$-model at varying inverse temperatures (a) $\beta=1.2$, (b) $\beta=0.9$, (c) $\beta=0.6$, and (d) $\beta=0.3$. All networks were created with the following parameters: $N=5000$, $\langle k\rangle = 12$, $\gamma = 2.5$. }
	\label{fig:thetavtheta}
\end{figure}
In Fig.~\ref{fig:thetavtheta} we show the performance of Mercator for various inverse temperatures $\beta$ and $\gamma=2.5$. We see that, as expected, the embedding gets progressively worse as the temperature increases. As $\beta\rightarrow 0$, the coupling to the geometry becomes weaker and there is thus less geometric information contained in the topology. We do, however, observe that the embedding is still good even for $\beta$'s relatively far below the transition point. For small values of $\beta \lesssim 0.6$, Mercator is not able to infer the correct $\beta$ because the geometric coupling in these networks is extremely weak. Even so, by feeding Mercator with the correct value of $\beta$, we are able to obtain an embedding. However, as shown in Fig~\ref{fig:thetavtheta}d, the obtained coordinates are completely different from the original ones so that, even if the nodes were originally placed on an underlying geometry, the resulting topology is not congruent with it. \\
Now that we have an intuitive idea of the performance of Mercator in the weakly geometric region $0<\beta<1$, the next step is to substantiate these results. To this end, we generate twenty network realizations for particular values of $\gamma$ and a range of $\beta$'s. We then embed these and test the quality of the embedding. In Fig.~\ref{fig:cscore} we show the results for the $C$-score, which measures how well the original ordering of the nodes on the circle is reproduced in the embedding (we provide a complete definition of the $C$-score in Appendix~\ref{sec:cscore}). We observe a transition between almost perfect reproduction of the ordering by Mercator ($C = 1$) at high $\beta$, to a situation where the ordering is completely random ($C=1/2$) at low $\beta$. This confirms the results we obtained in Fig.~\ref{fig:thetavtheta}. We also note that the region where the embedding is still faithful is larger for higher $\gamma$, and that the variance in the $C$-score explodes at $\beta=2/\gamma$, indicated by the dotted lines in Fig.~\ref{fig:cscore}. This is in agreement with the theoretical results in Ref.~\cite{vanderKolk2022}. There, $\beta=\beta_c'=2/\gamma$ marked the transition between slow, temperature dependent, decay of the average local clustering coefficient for $\beta>\beta_c'$ and a faster decay for $\beta<\beta_c'$, equivalent to the one observed in the soft configuration model~\cite{ColomerdeSimon2012}. These two regions were then coined the quasi-geometric and non-geometric regimes, respectively. The fact that we recover this transition here is a very profound result, as it confirms that the division of the region $\beta<1$ into these two sub-regions is not just theoretical in nature but has very real observable consequences. 

\subsection{Artificial networks. Greedy routing}
\begin{figure}[t]
	\centering
	\includegraphics[width=\columnwidth]{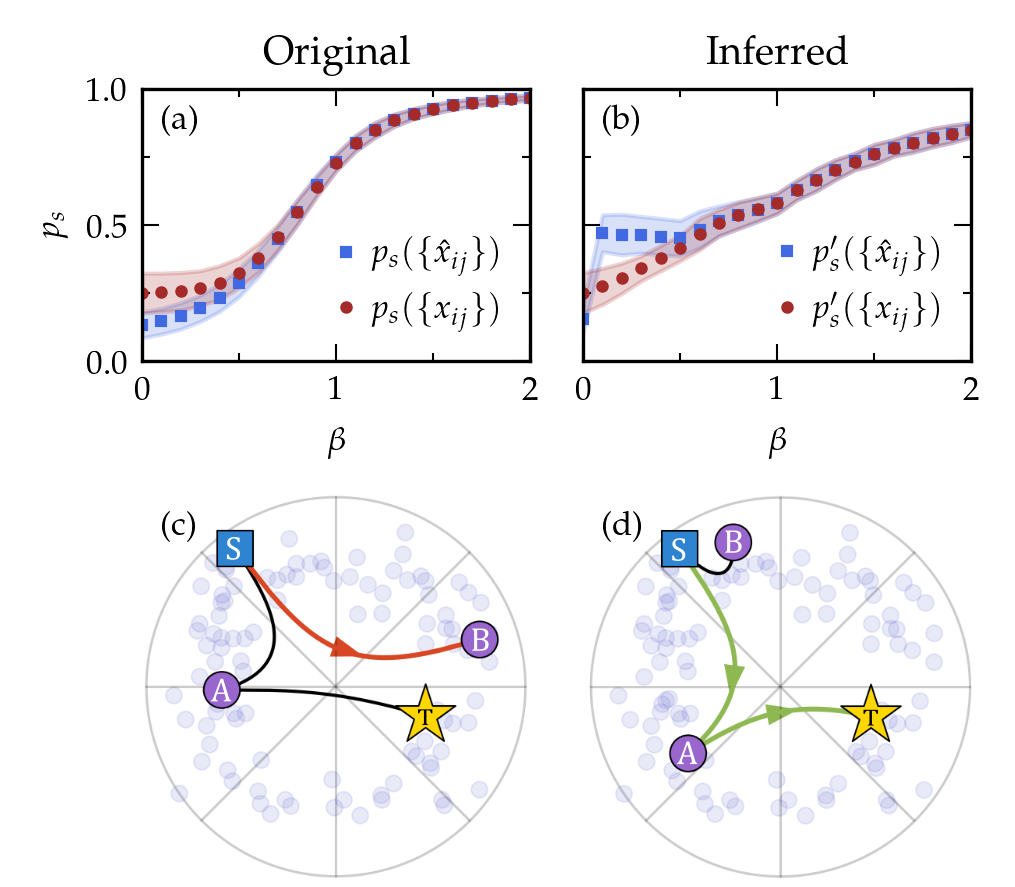}
	\caption{Results for the greedy routing routine. \textbf{(a)} Success probability $p_s$ based on the original coordinates, using both the definition of the hyperbolic distance where the contribution vanishes at $\beta=0$ ($x_{nm}$) as well as where it remains constant ($\hat{x}_{nm}$). The shaded regions represent the $2\sigma$ confidence interval. Similar results are shown in panel \textbf{(b)} where now the inferred coordinates are used. Parameters used: $\{N,\langle k\rangle,\gamma\}=\{4000,12,2.5\}$. In panels \textbf{(c)} and \textbf{(d)} a schematic representation of greedy routing paths based on the original and inferred coordinates, respectively, are shown. }
	\label{fig:greedyrouting}
\end{figure}

The second test we discuss here is more practical in nature and involves the performance of the greedy routing protocol~\cite{Krioukov2010}. In this protocol, a pair of nodes is selected at random, and the goal is to efficiently send a packet of information from one to the other. This is done by looking at the neighbors of the node that contains the packet, which is then forwarded to whichever neighbor is closer in hyperbolic space to the destination. This is repeated until one of two scenarios occurs. In scenario (1), the packet reaches the destination. In scenario (2), the neighbor closest to the goal is the node from which the parcel just arrived. In this latter case the packet is dropped as the destination cannot be reached using the greedy routing method.\\ 
One of the measures to define how well this algorithm performs is the success probability $p_s$, defined as the fraction of nodes pairs for which a greedy routing path exists. In Ref.~\cite{Krioukov2010} it was shown that information can be efficiently routed through the network if one uses the coordinates in the latent space. Of course, this works better when the connection to this underlying space is stronger, i.e. when $\beta$ is higher. This is confirmed in Fig.~\ref{fig:greedyrouting}a. When using the original coordinates and the hyperbolic distance as defined in Eq.~\ref{eq:hypdist_approx}, one observes that the success probability $p_s$ decays with $\beta$ until leveling out at $\beta=0$. Here, the angular coordinates are no longer taken into account and the greedy routing is purely based on the degrees of the network. If we redefine the hyperbolic distance such that it reads
\begin{equation}
	\hat{x}_{nm} = r_n + r_m + 2\ln\frac{\Delta\theta}{2}\label{eq:xij-2},
\end{equation}
i.e. such that the effect of the angular coordinates does not diminish, we see that the results are worse. This is because, for extremely low $\beta$, the connection between the topology and the geometry is lost and the angular coordinates are thus meaningless, impeding proper routing.

Let us now turn to the inferred coordinates, for which the results can be found in Fig.~\ref{fig:greedyrouting}b. Returning to the original definition of the hyperbolic distance, we see that for, $\beta<1$, the results are better than in the case of the original coordinates. This can be understood as follows: As lowering $\beta$ can be interpreted as increasing the temperature, more of the connectivity is determined by noise (conditioned on the hidden degrees). However, Mercator will always try and find as much geometry as possible, and place nodes in such a way that the inputted network realization is most congruent with it. In practice, this means that two nodes that were originally far away from each other, but are connected due to the large fluctuations, will most likely be placed close together in the embedding. In other words, the fact that for finite systems even non-geometric random graphs display clustering implies that an effective geometry can be found such that the effect of the triangle inequality on the topology is strongest. This is reminiscent of the fact that fluctuations in random graphs can lead to high modularity~\cite{Guimera2004}, which can lead to the detection of spurious communities~\cite{Peixoto2023}. In our case, Mercator is able to uncover an effective geometry, arising from the noise in the system (which makes this a finite size effect). However, where detecting spurious communities can be considered undesirable, the effective geometry can be useful. 
For example, it is beneficial to the greedy routing routine, as nodes that are close together are now also connected. Of course, when using the original hyperbolic distance, this effect will eventually disappear as the angular coordinates are no longer taken into account, leading to a success probability that coincides with that of the original coordinates at $\beta=0$. However, if we again use Eq.~\ref{eq:xij-2}, keeping the influence of $\theta$ constant, we note that high success probabilities can be achieved, even for low betas. Note that exactly at $\beta=0$ this is no longer the case as Mercator does not even try to find meaningful angular coordinates. \\
In Fig.~\ref{fig:greedyrouting}c,d this effect is clarified graphically. Here, the goal is to send a packet from the source node labeled $S$ to the target node labeled $T$. There is only one correct path, passing through node $A$. In the original metric space, due to fluctuations, the source node is also connected to node $B$, even though it lies far away from it. As $B$ lies closer to the target, the packet will get forwarded there. There is, however, no connection between $B$ and $T$ and so the packet gets dropped. In the case of the inferred coordinates, node $B$ gets placed closer to the source, and further from the target, in accordance with its connectivity. Node $A$ now lies closer to the target than $B$ does, and so a successful routing is achieved.\\
\subsection{Real Networks}
\begin{table*}[t]
	\caption{\label{tab:realnets} Set of real networks classified by type: ($\textrm{I}$) non-geometric, ($\textrm{II}$) quasi-geometric. Network properties are also shown. The following abbreviations are used: (MB) Metabolic, (GI) Genetic Interactions, (GMP) Genetic Multiplex, (PPI) Protein Protein Interactions, (PoP) Point of Presence. Detailed descriptions of the networks can be found in App.~\ref{app:realnets}.  }
	\begin{ruledtabular}
		\begin{tabular}{lccccccc}
			\textbf{Network} 			& \textbf{Area}& \textbf{$N$} & \textbf{$\langle k\rangle$} & $k_{\text{max}}$ &  \textbf{$\overline{c}$} & \textbf{$\beta$} & \textbf{Type}   \\\hline
			Foodweb--Eocene 				&Ecological		& $700$		&$18.3$		&$192$	&$0.10$&$\beta\approx0$&$\textrm{I}$\\
			Foodweb--Wetland			    &Ecological		& $128$ 	& $32.4$ 	&$110$	& $0.33$ & $\beta\approx0$ & $\textrm{I}$\\
			WordAdjacency--English 	    & Language 		&$7377$		&$12.0$		&$2568$	&$0.47$&$\beta\approx0$&$\textrm{I}$\\
			WordAdjacency--Japanese	    & Language		& $2698$ 	& $5.9$ 	&$725 $ 	& $0.30$ & $\beta\approx0$&$\textrm{I}$ \\
			MB--R.norvegicus			    &Cell 			& $1590$ 	& $5.9$ 	&$498 $ 	& $0.19$ & $\beta\approx0$&$\textrm{I}$ \\ 
			WikiTalk--Catalan 			&Social			& $79209$ 	& $4.6$ 	& $53234 $ 	& $0.83$ & $\beta\approx0$ & $\textrm{I}$\\ 
			GI--S.cerevisiae			    &Cell			&$5933$ 	& $149$ 	& $2244$ & $0.17$ & $0.63$ & $\textrm{II}$ \\ 
			GMP--C.elegans 				&Cell		  	& $3692$	&$4.1$		&$526$	&$0.11$&$0.69$&$\textrm{II}$\\
			Gnutella 					&Technological	& $10876$	& $7.4$ 	& $103$ & $0.01$ & $0.73$ & $\textrm{II}$ \\
			PPI--S.cerevisiae 			&Cell			& $7271$ 	& $45.0$ 	& $3613$ & $0.37$ & $0.75$ & $\textrm{II}$\\ 
			PPI--D.melanogaster 			&Cell 			& $11319$ 	& $23.7$ 	& $889 $ 	& $0.10$ & $0.84$ & $\textrm{II}$\\
			Transport--London		    &Transportation	& $369$ 	& $2.3$ 	&$7$	& $0.03$ & $0.86$ & $\textrm{II}$\\ 
			GMP--S.cerevisiae		    &Cell			& $6567$	&$68.1$		&$3254$	&$0.22$&$0.88$&$\textrm{II}$\\
			Internet-PoP 				&Technological	& $754$ 	& $2.4$ 	& $7$ & $0.03$ & $0.90$ & $\textrm{II}$\\
			PPI--H.sapiens				&Cell			&$27578$	&$37.9$		&$2883$ &$0.15$	&$0.91$ &$\textrm{II}$\\
			WikiVote				    &Social			& $7066$ 	& $28.5$ 	& $1065$ & $0.21$ & $0.91$ & $\textrm{II}$\\
			MathOverflow 				&Social			& $13599$ 	& $10.5$ 	& $949 $ 	& $0.32$ & $0.99$ &$\textrm{II}$\\
		\end{tabular}
	\end{ruledtabular}
\end{table*}
\begin{figure*}[t]
	\centering
	\includegraphics[width=2\columnwidth]{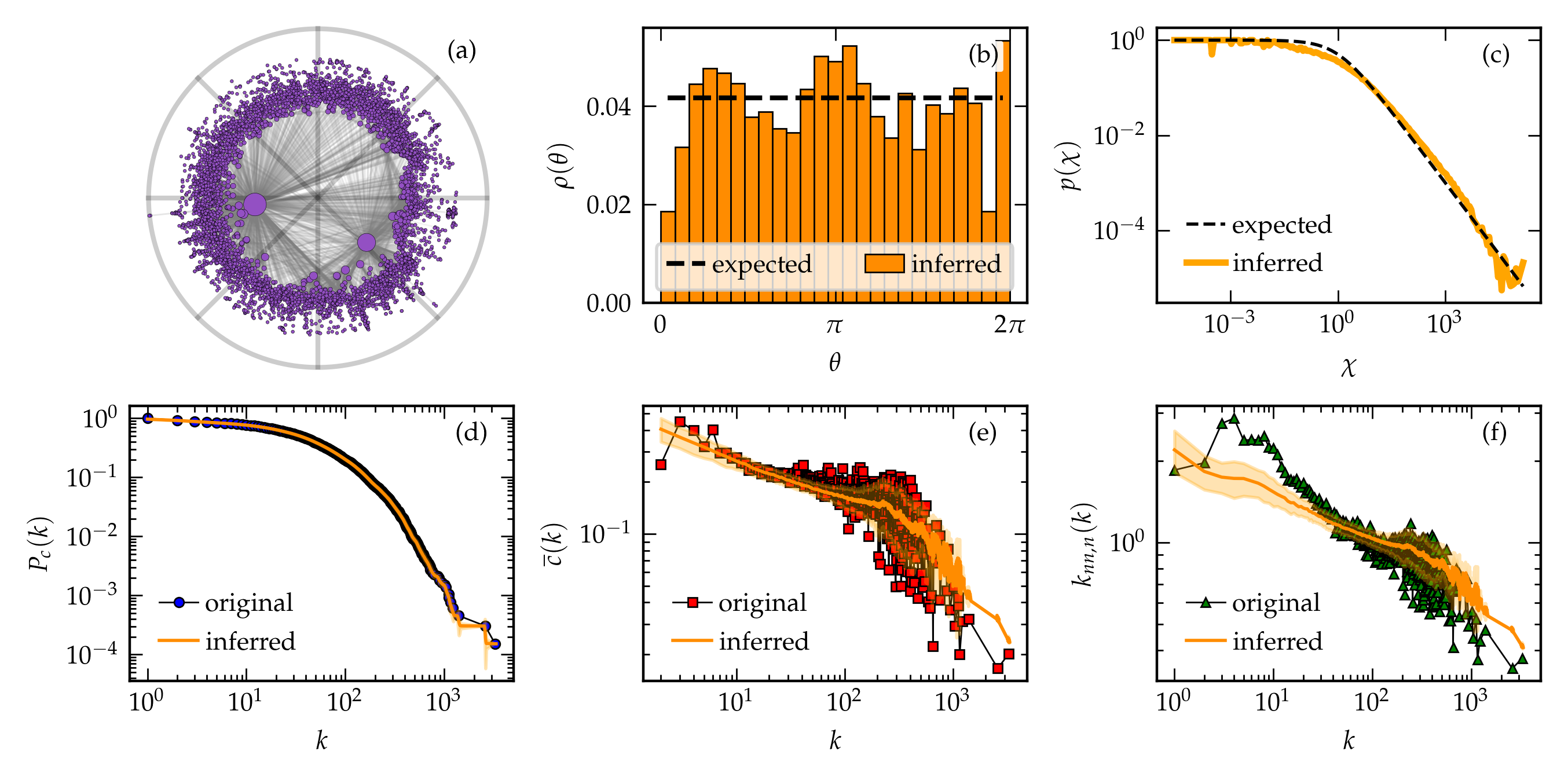}
	\caption{Summary of the results of Mercator for the GMP–S.cerevisiae network. \textbf{(a)} Representation of the embedding in the hyperbolic plane as defined by the $\mathbb{H}^2$-model. The top 5\% most geometric edges are shown. \textbf{(b)} Comparison between the expected and inferred densities of nodes along the circle. \textbf{(c)} Comparison between the probability distribution as expected based on the model (expected) as well as the actual distribution based on the inferred coordinates (inferred). The reproduction of the topological properties is also given: \textbf{(d)} the complementary cumulative degree distribution, \textbf{(e)} the average local clustering coefficient per degree class and \textbf{(f)} the degree-degree correlations per degree class. The inferred results are obtained by generating 100 realizations of the $\mathbb{S}^1$-model based on the inferred coordinates. The orange shaded regions represent the $2\sigma$ confidence interval.}
	\label{fig:realnet}
\end{figure*}
So far our study has been based on analyzing artificial networks generated from the $\mathbb{S}^1$-model. In the following, we show that many real networks are best described by the region $\beta<1$. In the case of real networks, the theoretical transition point $\beta'_c = 2/\gamma$ between the non- and quasi-geometric regions is less useful as it is mostly not possible to accurately extract the exponent $\gamma$. We therefore define a new classification of three distinct types of networks. Type $\mathrm{I}$ networks are classified by Mercator as being effectively non-geometric, i.e. their clustering can be explained by the configuration model. Type $\mathrm{II}$ networks live in the region $\beta<1$, but Mercator is still able to determine their temperature as they have significantly more clustering than one would expect from a network generated by the configuration model. These networks can be considered quasi-geometric. Finally, type $\mathrm{III}$ networks are those network that have $\beta>1$ and, thus, they are fully geometric.\\
In Tab.~\ref{tab:realnets} we show a selection of real networks and their properties, as well as their classification into the categories described above. There are several real networks for both categories $\textrm{I}$ and $\textrm{II}$, where $\beta<1$. Note also that the absolute value of the average local clustering is, on its own, not a good indicator for geometric coupling strength. For example, the value of $\overline{c}$ for the Words network is relatively high, but it is still a class $\textrm{I}$ network. This is because the triangles in the network can also be formed due to the presence of high degree nodes, meaning that this level of average local clustering can also be obtained in the configuration model. Another interesting observation is the presence of several gene regulatory and protein-protein interaction networks in the region $\beta<1$, as well as the fact that several ecological networks are deemed to have extremely weak geometry. 

In Fig.~\ref{fig:realnet} we show the full results for the embedding for the SCerevisiase-G network, the genetic multiplex network of the variety of yeast used in brewing beer. We see that the topological properties of the real network are reproduced very well by Mercator, even though the geometric coupling is weak.  
Similar results for the other real networks in Tab.~\ref{tab:realnets} can be found in the Supplementary Information~\cite{supp}. Interestingly, Mercator is still able to reproduce well the structural properties of class $\textrm{I}$ networks, where $\beta$ is set to zero manually. This does not come as a surprise, as these networks should be well described by the soft configuration model, i.e. on the basis of the hidden degrees alone. As Mercator is still able to set these such that the inferred degree sequence matches the observed one, the other structural properties and the profile of the connection probability are reproduced as well.

\section{Discussion}
Network geometry is an important framework that can explain many structural properties of real networks. In the past, the question if real networks are geometric was framed as being binary: They are or they are not. However, recent results~\cite{vanderKolk2022} indicate that the transition between these regimes is in fact not as hard for finite sized networks, and the existence of a quasi geometric-regime was conjectured. In this region, the clustering vanishes in the thermodynamic limit, implying that geometry is lost, but it does so very slowly as a function of the system size, which means that for real network, which are of finite size, clustering is significant. In this paper, we studied what this implies for the geometricity of a network. \\
We extended Mercator, an accurate and efficient tool for network embeddings based on the Popularity$\times$Similarity $\mathbb{S}^1$-model, to the weak coupling regime. We then showed that, in the quasi-geometric region, the tool is able to recover a significant amount of geometric information only from the topology of the network. This implies that the geometry is indeed relevant in this regime. Only when the coupling is very close to zero does the geometricity completely vanish. Here, the properties of the network can be explained by the soft configuration model. On the basis of these findings we define three classes of networks: (class-I) Those with weak geometric coupling where the topology is completely defined by the degrees, (class-II) those with weak geometric coupling but where the similarity dimension also plays a role in defining the topology and (class-III) those with strong geometric coupling where the similarity dimensions is very important. We show that real networks are represented in all three of these categories, which means that it is important to take them all into account when discussing the geometry of real networks. We show that the presence of triangles in finite non-geometric random graphs allows for the definition of an effective geometry by Mercator. This geometry does, therefore, not reflect the original underlying geometry, which is absent, but can still be useful for information routing problems. Future work might investigate if this effective geometry can also be used in other tasks, for example link prediction.  \\
All in all, this paper shows that the discussion of whether networks are geometric or not is limiting because the transition between these two extremes is not abrupt for finite sized networks. Indeed, many real networks are better described as being quasi-geometric, and even a weak geometric coupling can imply that geometric information is stored in the topology of a network. 

\acknowledgements
We acknowledge support from: Grant TED2021-129791B-I00 funded by MCIN/AEI/10.13039/501100011033 and the  ``European Union NextGenerationEU/PRTR''; Grants PID2019-106290GB-C22 and PID2022-137505NB-C22 funded by MCIN/AEI/10.13039/501100011033; Generalitat de Catalunya grant number 2021SGR00856. M. B. acknowledges the ICREA Academia award, funded by the Generalitat de Catalunya. J.~vd~K. acknowledges support from the Ministry of Universities of Spain in the form of the FPU predoctoral contract.

\appendix

\section{C-score}\label{sec:cscore}
The concordance or C-score is a measure that quantifies the concordance between two different orderings. In our case, the first ordering is given by the set of vertices in a network, ordered by their original coordinates and the second by ordering the indices according to the inferred coordinates. First introduced in Ref.~\cite{Zagar2011}, the C-score was adjusted to a system with periodic boundary conditions in Ref.~\cite{Muscoloni2017}, leading to the following definition:
\begin{equation}
	\text{C-score} = \frac{2}{N(N-1)}\sum\limits_{n=1}^{N-1}\sum\limits_{m=n+1}^N \delta(n,m),
\end{equation}
where $N$ is the total amount of nodes, $n$ and $m$ indicate two nodes and $\delta(n,m)$ is $1$ if the shortest distance between $n$ and $m$ along the circle has the same direction (clockwise or counterclockwise) in both the original and inferred ordering, and $0$ if the direction is different. Note that it is possible that Mercator returns an inverted ordering, which, for example, leads to an inverted diagonal in Fig.~\ref{fig:thetavtheta}, as well as a $C$-score $<0.5$. Of course, the orientation of the ring does not influence the quality of the embedding, as it is only the distance between points along the circle that matters. Therefore, we are actually interested in using $\max{(\text{C-score},1-(\text{C-score}))}$ as a measure, such that $1$ implies perfect ordering and $0.5$ means the inferred order is completely random.

\section{Real Networks}\label{app:realnets}
In this section we give an overview of the networks studied in the main text (Tab.~\ref{tab:realnets}).

\begin{itemize}
	
	\item \textbf{Foodweb-Eocene}~\cite{Dunne2014}: A reconstructed food web of an ecosystem from the early Eocene (48 million years ago). Nodes represent taxa and edges represent consumer-resource relations. The original network was directed.  
	\item \textbf{Foodweb-Wetland}~\cite{Ulanowicz2005}: A network of carbon exchanges among species in the cypress wetlands of South Florida. Nodes represent taxa and edges represent consumer-resource relations. The original network was directed. 
	\item \textbf{WordAdjacency-English}~\cite{Milo2004}: A network of word adjacency in English texts. Nodes represent words and two words are connected if one directly follows the other in texts The original network was directed. 
	\item \textbf{WordAdjacency-Japanese}~\cite{Milo2004}: A network of word adjacency in Japanese texts. Nodes represent words and two words are connected if one directly follows the other in texts The original network was directed.
	\item \textbf{MB-R.norvegicus}~\cite{Huss2007}: A metabolic network of the rat (Ratus norvegicus), extracted from the Kyoto Encyclopedia of Genes and Genomes (KEGG). Nodes represent substances involved in enzymatic reactions and edges represent reactant-product pairs. 
	\item \textbf{WikiTalk-Catalan}~\cite{Kunegis2013}: A network where nodes represents Wikipedia editors for a certain language (in this case Catalan), and where user $i$ and $j$ are connected if $i$ leaves a message on the talk page of $j$. The original network was directed. 
	\item \textbf{GI-S.cerevisiae}~\cite{Hu2018}: A network based on the Molecular Interaction Search Tool (MIST) for baker's yeast (Saccharomyces cerevisiae). Here node represent genes and the edges indicate that the effects of mutations in one gene can be modified by mutations of another gene. 
	\item \textbf{GMP-C.elegans}~\cite{DeDomenico2015}: A multiplex network representing different types of genetic interactions for the nematode worm Caenorhabditis elegans. The layers represent physical, association, co-localization, direct, suppressive and additive interactions. In this paper we create a monolayer network by treating the different interaction types equally and removing double links. The original network was directed. 
	\item \textbf{Gnutella}~\cite{Ripeanu2002}: A snapshot of the Gnutella peer-to-peer file sharing network on August 4th 2002. Nodes are hosts and edges are connections between them. The original network was directed. 
	\item  \textbf{PPI-S.cerevisiae}~\cite{Hu2018}: A network based on the Molecular Interaction Search Tool (MIST) for baker's yeast (Saccharomyces cerevisiae). Here node represent genes and the edges indicate that there are physical interactions between their associated proteins. 
	\item \textbf{PPI-D.melanogaster}~\cite{Hu2018}: A network based on the Molecular Interaction Search Tool (MIST) for the fruit fly (Drosophila melanogaster). Here node represent genes and the edges indicate that there are physical interactions between their associated proteins.
	\item \textbf{Transport-London}~\cite{DeDomenico2014}: An multiplex network of the public transportation system in London. Nodes are London train stations and the links can represent either the underground, overground and DLR connections. There connections are treated equally as to create a mono-layer network. 
	\item \textbf{GMP-S.cerevisiae}~\cite{DeDomenico2015}: A multiplex network representing different types of genetic interactions for baker's yeast (Saccharomyces cerevisiae). The layers represent physical, association, co-localization, direct, suppressive and additive interactions. In this paper we create a monolayer network by treating the different interaction types equally and removing double links. The original network was directed. 
	\item \textbf{Internet-PoP}~\cite{Knight2011}: The Kentucky Datalink network, an internet graph at the Point of Presence (PoP) level. NODES AND EDGES
	\item \textbf{PPI-H.sapiens}~\cite{Hu2018}: A network based on the Molecular Interaction Search Tool (MIST) for humans (Homo sapiens). Here node represent genes and the edges indicate that there are physical interactions between their associated proteins.
	\item \textbf{WikiVote}~\cite{Leskovec2010}: The network represents the voting process used to select Wikipedia administrators, which are contributors with access to additional technical features. Nodes represents Wikipedia users and an edge is created if user $i$ votes on the selections of user $j$. The original network was directed.
	\item \textbf{MathOverflow}~\cite{Paranjape2017}: An interaction network of users (nodes) on the online Q\&A site MathOverflow. An edge from node $i$ to node $j$ indicates that $i$ responded to an answer by $j$. The original network was directed. 
\end{itemize}

\bibliography{bibliography}

\begin{thebibliography}{63}%
\makeatletter
\providecommand \@ifxundefined [1]{%
 \@ifx{#1\undefined}
}%
\providecommand \@ifnum [1]{%
 \ifnum #1\expandafter \@firstoftwo
 \else \expandafter \@secondoftwo
 \fi
}%
\providecommand \@ifx [1]{%
 \ifx #1\expandafter \@firstoftwo
 \else \expandafter \@secondoftwo
 \fi
}%
\providecommand \natexlab [1]{#1}%
\providecommand \enquote  [1]{``#1''}%
\providecommand \bibnamefont  [1]{#1}%
\providecommand \bibfnamefont [1]{#1}%
\providecommand \citenamefont [1]{#1}%
\providecommand \href@noop [0]{\@secondoftwo}%
\providecommand \href [0]{\begingroup \@sanitize@url \@href}%
\providecommand \@href[1]{\@@startlink{#1}\@@href}%
\providecommand \@@href[1]{\endgroup#1\@@endlink}%
\providecommand \@sanitize@url [0]{\catcode `\\12\catcode `\$12\catcode
  `\&12\catcode `\#12\catcode `\^12\catcode `\_12\catcode `\%12\relax}%
\providecommand \@@startlink[1]{}%
\providecommand \@@endlink[0]{}%
\providecommand \url  [0]{\begingroup\@sanitize@url \@url }%
\providecommand \@url [1]{\endgroup\@href {#1}{\urlprefix }}%
\providecommand \urlprefix  [0]{URL }%
\providecommand \Eprint [0]{\href }%
\providecommand \doibase [0]{https://doi.org/}%
\providecommand \selectlanguage [0]{\@gobble}%
\providecommand \bibinfo  [0]{\@secondoftwo}%
\providecommand \bibfield  [0]{\@secondoftwo}%
\providecommand \translation [1]{[#1]}%
\providecommand \BibitemOpen [0]{}%
\providecommand \bibitemStop [0]{}%
\providecommand \bibitemNoStop [0]{.\EOS\space}%
\providecommand \EOS [0]{\spacefactor3000\relax}%
\providecommand \BibitemShut  [1]{\csname bibitem#1\endcsname}%
\let\auto@bib@innerbib\@empty
\bibitem [{\citenamefont {Dunne}(2005)}]{Dunne2005}%
  \BibitemOpen
  \bibfield  {author} {\bibinfo {author} {\bibfnamefont {J.}~\bibnamefont
  {Dunne}},\ }\href@noop {} {\bibinfo {title} {The network structure of food
  webs}} (\bibinfo {year} {2005})\BibitemShut {NoStop}%
\bibitem [{\citenamefont {Sporns}(2013)}]{Sporns2013}%
  \BibitemOpen
  \bibfield  {author} {\bibinfo {author} {\bibfnamefont {O.}~\bibnamefont
  {Sporns}},\ }\bibfield  {title} {\bibinfo {title} {Structure and function of
  complex brain networks},\ }\href
  {https://doi.org/10.31887/DCNS.2013.15.3/osporns} {\bibfield  {journal}
  {\bibinfo  {journal} {Dialogues in Clinical Neuroscience}\ }\textbf {\bibinfo
  {volume} {15}},\ \bibinfo {pages} {247} (\bibinfo {year} {2013})}\BibitemShut
  {NoStop}%
\bibitem [{\citenamefont {Freeman}(2004)}]{Freeman2004}%
  \BibitemOpen
  \bibfield  {author} {\bibinfo {author} {\bibfnamefont {L.~C.}\ \bibnamefont
  {Freeman}},\ }\href@noop {} {\emph {\bibinfo {title} {The Development of
  Social Network Analysis}}}\ (\bibinfo  {publisher} {Empirical Press},\
  \bibinfo {year} {2004})\BibitemShut {NoStop}%
\bibitem [{\citenamefont {Barab{\'a}si}(2001)}]{Barabasi2001}%
  \BibitemOpen
  \bibfield  {author} {\bibinfo {author} {\bibfnamefont {A.-L.}\ \bibnamefont
  {Barab{\'a}si}},\ }\bibfield  {title} {\bibinfo {title} {The physics of the
  web},\ }\href {https://doi.org/10.1088/2058-7058/14/7/32} {\bibfield
  {journal} {\bibinfo  {journal} {Physics World}\ }\textbf {\bibinfo {volume}
  {14}},\ \bibinfo {pages} {33} (\bibinfo {year} {2001})}\BibitemShut {NoStop}%
\bibitem [{\citenamefont {Buchanan}(2003)}]{Buchanan2003}%
  \BibitemOpen
  \bibfield  {author} {\bibinfo {author} {\bibfnamefont {M.}~\bibnamefont
  {Buchanan}},\ }\href@noop {} {\emph {\bibinfo {title} {Nexus: Small Worlds
  and the Groundbreaking Theory of Networks}}}\ (\bibinfo  {publisher} {W.W.
  Norton and Company},\ \bibinfo {year} {2003})\BibitemShut {NoStop}%
\bibitem [{\citenamefont {Albert}\ and\ \citenamefont
  {Barab{\'a}si}(2002)}]{Albert2002}%
  \BibitemOpen
  \bibfield  {author} {\bibinfo {author} {\bibfnamefont {R.}~\bibnamefont
  {Albert}}\ and\ \bibinfo {author} {\bibfnamefont {A.-L.}\ \bibnamefont
  {Barab{\'a}si}},\ }\bibfield  {title} {\bibinfo {title} {Statistical
  mechanics of complex networks},\ }\href
  {https://doi.org/10.1103/RevModPhys.74.47} {\bibfield  {journal} {\bibinfo
  {journal} {Reviews of Modern Physics}\ }\textbf {\bibinfo {volume} {74}},\
  \bibinfo {pages} {47} (\bibinfo {year} {2002})}\BibitemShut {NoStop}%
\bibitem [{\citenamefont {Newman}(2003)}]{Newman2003}%
  \BibitemOpen
  \bibfield  {author} {\bibinfo {author} {\bibfnamefont {M.~E.~J.}\
  \bibnamefont {Newman}},\ }\bibfield  {title} {\bibinfo {title} {The structure
  and function of complex networks},\ }\href
  {https://doi.org/10.1137/S003614450342480} {\bibfield  {journal} {\bibinfo
  {journal} {SIAM Review}\ }\textbf {\bibinfo {volume} {45}},\ \bibinfo {pages}
  {167} (\bibinfo {year} {2003})}\BibitemShut {NoStop}%
\bibitem [{\citenamefont {Bogu{\~n}{\'a}}\ \emph {et~al.}(2021)\citenamefont
  {Bogu{\~n}{\'a}}, \citenamefont {Bonamassa}, \citenamefont {Domenico},
  \citenamefont {Havlin}, \citenamefont {Krioukov},\ and\ \citenamefont
  {Serrano}}]{Boguna2021}%
  \BibitemOpen
  \bibfield  {author} {\bibinfo {author} {\bibfnamefont {M.}~\bibnamefont
  {Bogu{\~n}{\'a}}}, \bibinfo {author} {\bibfnamefont {I.}~\bibnamefont
  {Bonamassa}}, \bibinfo {author} {\bibfnamefont {M.~D.}\ \bibnamefont
  {Domenico}}, \bibinfo {author} {\bibfnamefont {S.}~\bibnamefont {Havlin}},
  \bibinfo {author} {\bibfnamefont {D.}~\bibnamefont {Krioukov}},\ and\
  \bibinfo {author} {\bibfnamefont {M.~{\'A}.}\ \bibnamefont {Serrano}},\
  }\bibfield  {title} {\bibinfo {title} {Network geometry},\ }\href
  {https://doi.org/10.1038/s42254-020-00264-4} {\bibfield  {journal} {\bibinfo
  {journal} {Nature Reviews Physics}\ }\textbf {\bibinfo {volume} {3}},\
  \bibinfo {pages} {114} (\bibinfo {year} {2021})}\BibitemShut {NoStop}%
\bibitem [{\citenamefont {Serrano}\ \emph {et~al.}(2008)\citenamefont
  {Serrano}, \citenamefont {Krioukov},\ and\ \citenamefont
  {Bogu{\~n}{\'a}}}]{Serrano2008}%
  \BibitemOpen
  \bibfield  {author} {\bibinfo {author} {\bibfnamefont {M.~{\'A}.}\
  \bibnamefont {Serrano}}, \bibinfo {author} {\bibfnamefont {D.}~\bibnamefont
  {Krioukov}},\ and\ \bibinfo {author} {\bibfnamefont {M.}~\bibnamefont
  {Bogu{\~n}{\'a}}},\ }\bibfield  {title} {\bibinfo {title} {Self-similarity of
  complex networks and hidden metric spaces},\ }\href
  {https://doi.org/10.1103/PhysRevLett.100.078701} {\bibfield  {journal}
  {\bibinfo  {journal} {Physical Review Letters}\ }\textbf {\bibinfo {volume}
  {100}},\ \bibinfo {pages} {078701} (\bibinfo {year} {2008})}\BibitemShut
  {NoStop}%
\bibitem [{\citenamefont {M{\"u}ller}\ and\ \citenamefont
  {Staps}(2019)}]{Muller2019}%
  \BibitemOpen
  \bibfield  {author} {\bibinfo {author} {\bibfnamefont {T.}~\bibnamefont
  {M{\"u}ller}}\ and\ \bibinfo {author} {\bibfnamefont {M.}~\bibnamefont
  {Staps}},\ }\bibfield  {title} {\bibinfo {title} {The diameter of kpkvb
  random graphs},\ }\href {https://doi.org/10.1017/apr.2019.23} {\bibfield
  {journal} {\bibinfo  {journal} {Advances in Applied Probability}\ }\textbf
  {\bibinfo {volume} {51}},\ \bibinfo {pages} {358} (\bibinfo {year}
  {2019})}\BibitemShut {NoStop}%
\bibitem [{\citenamefont {Friedrich}\ and\ \citenamefont
  {Krohmer}(2018)}]{Friedrich2018}%
  \BibitemOpen
  \bibfield  {author} {\bibinfo {author} {\bibfnamefont {T.}~\bibnamefont
  {Friedrich}}\ and\ \bibinfo {author} {\bibfnamefont {A.}~\bibnamefont
  {Krohmer}},\ }\bibfield  {title} {\bibinfo {title} {On the diameter of
  hyperbolic random graphs},\ }\href {https://doi.org/10.1137/17M1123961}
  {\bibfield  {journal} {\bibinfo  {journal} {SIAM Journal on Discrete
  Mathematics}\ }\textbf {\bibinfo {volume} {32}},\ \bibinfo {pages} {1314}
  (\bibinfo {year} {2018})}\BibitemShut {NoStop}%
\bibitem [{\citenamefont {Fountoulakis}\ \emph {et~al.}(2021)\citenamefont
  {Fountoulakis}, \citenamefont {van~der Hoorn}, \citenamefont {M{\"u}ller},\
  and\ \citenamefont {Schepers}}]{Fountoulakis2021}%
  \BibitemOpen
  \bibfield  {author} {\bibinfo {author} {\bibfnamefont {N.}~\bibnamefont
  {Fountoulakis}}, \bibinfo {author} {\bibfnamefont {P.}~\bibnamefont {van~der
  Hoorn}}, \bibinfo {author} {\bibfnamefont {T.}~\bibnamefont {M{\"u}ller}},\
  and\ \bibinfo {author} {\bibfnamefont {M.}~\bibnamefont {Schepers}},\
  }\bibfield  {title} {\bibinfo {title} {Clustering in a hyperbolic model of
  complex networks},\ }\bibfield  {journal} {\bibinfo  {journal} {Electronic
  Journal of Probability}\ }\textbf {\bibinfo {volume} {26}},\ \href
  {https://doi.org/10.1214/21-EJP583} {10.1214/21-EJP583} (\bibinfo {year}
  {2021})\BibitemShut {NoStop}%
\bibitem [{\citenamefont {Candellero}\ and\ \citenamefont
  {Fountoulakis}(2016)}]{Candellero2016}%
  \BibitemOpen
  \bibfield  {author} {\bibinfo {author} {\bibfnamefont {E.}~\bibnamefont
  {Candellero}}\ and\ \bibinfo {author} {\bibfnamefont {N.}~\bibnamefont
  {Fountoulakis}},\ }\bibfield  {title} {\bibinfo {title} {Clustering and the
  hyperbolic geometry of complex networks},\ }\href
  {https://doi.org/10.1080/15427951.2015.1067848} {\bibfield  {journal}
  {\bibinfo  {journal} {Internet Mathematics}\ }\textbf {\bibinfo {volume}
  {12}},\ \bibinfo {pages} {2} (\bibinfo {year} {2016})}\BibitemShut {NoStop}%
\bibitem [{\citenamefont {Gugelmann}\ \emph {et~al.}(2012)\citenamefont
  {Gugelmann}, \citenamefont {Panagiotou},\ and\ \citenamefont
  {Peter}}]{Gugelmann2012}%
  \BibitemOpen
  \bibfield  {author} {\bibinfo {author} {\bibfnamefont {L.}~\bibnamefont
  {Gugelmann}}, \bibinfo {author} {\bibfnamefont {K.}~\bibnamefont
  {Panagiotou}},\ and\ \bibinfo {author} {\bibfnamefont {U.}~\bibnamefont
  {Peter}},\ }\href {https://doi.org/10.1007/978-3-642-31585-5_51} {\bibinfo
  {title} {Random hyperbolic graphs: Degree sequence and clustering}} (\bibinfo
  {year} {2012})\BibitemShut {NoStop}%
\bibitem [{\citenamefont {Krioukov}\ \emph {et~al.}(2010)\citenamefont
  {Krioukov}, \citenamefont {Papadopoulos}, \citenamefont {Kitsak},
  \citenamefont {Vahdat},\ and\ \citenamefont {Bogu{\~n}{\'a}}}]{Krioukov2010}%
  \BibitemOpen
  \bibfield  {author} {\bibinfo {author} {\bibfnamefont {D.}~\bibnamefont
  {Krioukov}}, \bibinfo {author} {\bibfnamefont {F.}~\bibnamefont
  {Papadopoulos}}, \bibinfo {author} {\bibfnamefont {M.}~\bibnamefont
  {Kitsak}}, \bibinfo {author} {\bibfnamefont {A.}~\bibnamefont {Vahdat}},\
  and\ \bibinfo {author} {\bibfnamefont {M.}~\bibnamefont {Bogu{\~n}{\'a}}},\
  }\bibfield  {title} {\bibinfo {title} {Hyperbolic geometry of complex
  networks},\ }\href {https://doi.org/10.1103/PhysRevE.82.036106} {\bibfield
  {journal} {\bibinfo  {journal} {Physical Review E}\ }\textbf {\bibinfo
  {volume} {82}},\ \bibinfo {pages} {036106} (\bibinfo {year}
  {2010})}\BibitemShut {NoStop}%
\bibitem [{\citenamefont {Serrano}\ \emph {et~al.}(2012)\citenamefont
  {Serrano}, \citenamefont {Bogu{\~n}{\'a}},\ and\ \citenamefont
  {Sagu{\'e}s}}]{Serrano2012}%
  \BibitemOpen
  \bibfield  {author} {\bibinfo {author} {\bibfnamefont {M.~{\'A}.}\
  \bibnamefont {Serrano}}, \bibinfo {author} {\bibfnamefont {M.}~\bibnamefont
  {Bogu{\~n}{\'a}}},\ and\ \bibinfo {author} {\bibfnamefont {F.}~\bibnamefont
  {Sagu{\'e}s}},\ }\bibfield  {title} {\bibinfo {title} {Uncovering the hidden
  geometry behind metabolic networks},\ }\href
  {https://doi.org/10.1039/c2mb05306c} {\bibfield  {journal} {\bibinfo
  {journal} {Molecular BioSystems}\ }\textbf {\bibinfo {volume} {8}},\ \bibinfo
  {pages} {843} (\bibinfo {year} {2012})}\BibitemShut {NoStop}%
\bibitem [{\citenamefont {Garc{\'\i}a-P{\'e}rez}\ \emph
  {et~al.}(2016)\citenamefont {Garc{\'\i}a-P{\'e}rez}, \citenamefont
  {Bogu{\~n}{\'a}}, \citenamefont {Allard},\ and\ \citenamefont
  {Serrano}}]{GarciaPerez2016}%
  \BibitemOpen
  \bibfield  {author} {\bibinfo {author} {\bibfnamefont {G.}~\bibnamefont
  {Garc{\'\i}a-P{\'e}rez}}, \bibinfo {author} {\bibfnamefont {M.}~\bibnamefont
  {Bogu{\~n}{\'a}}}, \bibinfo {author} {\bibfnamefont {A.}~\bibnamefont
  {Allard}},\ and\ \bibinfo {author} {\bibfnamefont {M.~{\'A}.}\ \bibnamefont
  {Serrano}},\ }\bibfield  {title} {\bibinfo {title} {The hidden hyperbolic
  geometry of international trade: World trade atlas 1870--2013},\ }\href
  {https://doi.org/10.1038/srep33441} {\bibfield  {journal} {\bibinfo
  {journal} {Scientific Reports}\ }\textbf {\bibinfo {volume} {6}},\ \bibinfo
  {pages} {33441} (\bibinfo {year} {2016})}\BibitemShut {NoStop}%
\bibitem [{\citenamefont {Allard}\ and\ \citenamefont
  {Serrano}(2020)}]{Allard2020}%
  \BibitemOpen
  \bibfield  {author} {\bibinfo {author} {\bibfnamefont {A.}~\bibnamefont
  {Allard}}\ and\ \bibinfo {author} {\bibfnamefont {M.~{\'A}.}\ \bibnamefont
  {Serrano}},\ }\bibfield  {title} {\bibinfo {title} {Navigable maps of
  structural brain networks across species},\ }\href
  {https://doi.org/10.1371/journal.pcbi.1007584} {\bibfield  {journal}
  {\bibinfo  {journal} {PLOS Computational Biology}\ }\textbf {\bibinfo
  {volume} {16}},\ \bibinfo {pages} {e1007584} (\bibinfo {year}
  {2020})}\BibitemShut {NoStop}%
\bibitem [{\citenamefont {Guly{\'a}s}\ \emph {et~al.}(2015)\citenamefont
  {Guly{\'a}s}, \citenamefont {B{\'\i}r{\'o}}, \citenamefont {K{\H o}r{\"o}si},
  \citenamefont {R{\'e}tv{\'a}ri},\ and\ \citenamefont
  {Krioukov}}]{Gulyas2015}%
  \BibitemOpen
  \bibfield  {author} {\bibinfo {author} {\bibfnamefont {A.}~\bibnamefont
  {Guly{\'a}s}}, \bibinfo {author} {\bibfnamefont {J.~J.}\ \bibnamefont
  {B{\'\i}r{\'o}}}, \bibinfo {author} {\bibfnamefont {A.}~\bibnamefont {K{\H
  o}r{\"o}si}}, \bibinfo {author} {\bibfnamefont {G.}~\bibnamefont
  {R{\'e}tv{\'a}ri}},\ and\ \bibinfo {author} {\bibfnamefont {D.}~\bibnamefont
  {Krioukov}},\ }\bibfield  {title} {\bibinfo {title} {Navigable networks as
  nash equilibria of navigation games},\ }\href
  {https://doi.org/10.1038/ncomms8651} {\bibfield  {journal} {\bibinfo
  {journal} {Nature Communications}\ }\textbf {\bibinfo {volume} {6}},\
  \bibinfo {pages} {7651} (\bibinfo {year} {2015})}\BibitemShut {NoStop}%
\bibitem [{\citenamefont {Bogu{\~n}{\'a}}\ \emph {et~al.}(2009)\citenamefont
  {Bogu{\~n}{\'a}}, \citenamefont {Krioukov},\ and\ \citenamefont
  {Claffy}}]{Boguna2009}%
  \BibitemOpen
  \bibfield  {author} {\bibinfo {author} {\bibfnamefont {M.}~\bibnamefont
  {Bogu{\~n}{\'a}}}, \bibinfo {author} {\bibfnamefont {D.}~\bibnamefont
  {Krioukov}},\ and\ \bibinfo {author} {\bibfnamefont {K.~C.}\ \bibnamefont
  {Claffy}},\ }\bibfield  {title} {\bibinfo {title} {Navigability of complex
  networks},\ }\href {https://doi.org/10.1038/nphys1130} {\bibfield  {journal}
  {\bibinfo  {journal} {Nature Physics}\ }\textbf {\bibinfo {volume} {5}},\
  \bibinfo {pages} {74} (\bibinfo {year} {2009})}\BibitemShut {NoStop}%
\bibitem [{\citenamefont {Wang}\ \emph {et~al.}(2016)\citenamefont {Wang},
  \citenamefont {Wu}, \citenamefont {Li}, \citenamefont {Jin},\ and\
  \citenamefont {Xiong}}]{Wang2016}%
  \BibitemOpen
  \bibfield  {author} {\bibinfo {author} {\bibfnamefont {Z.}~\bibnamefont
  {Wang}}, \bibinfo {author} {\bibfnamefont {Y.}~\bibnamefont {Wu}}, \bibinfo
  {author} {\bibfnamefont {Q.}~\bibnamefont {Li}}, \bibinfo {author}
  {\bibfnamefont {F.}~\bibnamefont {Jin}},\ and\ \bibinfo {author}
  {\bibfnamefont {W.}~\bibnamefont {Xiong}},\ }\bibfield  {title} {\bibinfo
  {title} {Link prediction based on hyperbolic mapping with community structure
  for complex networks},\ }\href {https://doi.org/10.1016/j.physa.2016.01.010}
  {\bibfield  {journal} {\bibinfo  {journal} {Physica A: Statistical Mechanics
  and its Applications}\ }\textbf {\bibinfo {volume} {450}},\ \bibinfo {pages}
  {609} (\bibinfo {year} {2016})}\BibitemShut {NoStop}%
\bibitem [{\citenamefont {Kitsak}\ \emph {et~al.}(2020)\citenamefont {Kitsak},
  \citenamefont {Voitalov},\ and\ \citenamefont {Krioukov}}]{Kitsak2020}%
  \BibitemOpen
  \bibfield  {author} {\bibinfo {author} {\bibfnamefont {M.}~\bibnamefont
  {Kitsak}}, \bibinfo {author} {\bibfnamefont {I.}~\bibnamefont {Voitalov}},\
  and\ \bibinfo {author} {\bibfnamefont {D.}~\bibnamefont {Krioukov}},\
  }\bibfield  {title} {\bibinfo {title} {Link prediction with hyperbolic
  geometry},\ }\href {https://doi.org/10.1103/PhysRevResearch.2.043113}
  {\bibfield  {journal} {\bibinfo  {journal} {Physical Review Research}\
  }\textbf {\bibinfo {volume} {2}},\ \bibinfo {pages} {043113} (\bibinfo {year}
  {2020})}\BibitemShut {NoStop}%
\bibitem [{\citenamefont {Krioukov}(2016)}]{Krioukov2016}%
  \BibitemOpen
  \bibfield  {author} {\bibinfo {author} {\bibfnamefont {D.}~\bibnamefont
  {Krioukov}},\ }\bibfield  {title} {\bibinfo {title} {Clustering implies
  geometry in networks},\ }\href
  {https://doi.org/10.1103/PhysRevLett.116.208302} {\bibfield  {journal}
  {\bibinfo  {journal} {Physical Review Letters}\ }\textbf {\bibinfo {volume}
  {116}},\ \bibinfo {pages} {208302} (\bibinfo {year} {2016})}\BibitemShut
  {NoStop}%
\bibitem [{\citenamefont {Michielan}\ \emph {et~al.}(2022)\citenamefont
  {Michielan}, \citenamefont {Litvak},\ and\ \citenamefont
  {Stegehuis}}]{Michielan2022}%
  \BibitemOpen
  \bibfield  {author} {\bibinfo {author} {\bibfnamefont {R.}~\bibnamefont
  {Michielan}}, \bibinfo {author} {\bibfnamefont {N.}~\bibnamefont {Litvak}},\
  and\ \bibinfo {author} {\bibfnamefont {C.}~\bibnamefont {Stegehuis}},\
  }\bibfield  {title} {\bibinfo {title} {Detecting hyperbolic geometry in
  networks: Why triangles are not enough},\ }\href
  {https://doi.org/10.1103/PhysRevE.106.054303} {\bibfield  {journal} {\bibinfo
   {journal} {Physical Review E}\ }\textbf {\bibinfo {volume} {106}},\ \bibinfo
  {pages} {054303} (\bibinfo {year} {2022})}\BibitemShut {NoStop}%
\bibitem [{\citenamefont {van~der Kolk}\ \emph {et~al.}(2022)\citenamefont
  {van~der Kolk}, \citenamefont {Serrano},\ and\ \citenamefont
  {Bogu{\~n}{\'a}}}]{vanderKolk2022}%
  \BibitemOpen
  \bibfield  {author} {\bibinfo {author} {\bibfnamefont {J.}~\bibnamefont
  {van~der Kolk}}, \bibinfo {author} {\bibfnamefont {M.~{\'A}.}\ \bibnamefont
  {Serrano}},\ and\ \bibinfo {author} {\bibfnamefont {M.}~\bibnamefont
  {Bogu{\~n}{\'a}}},\ }\bibfield  {title} {\bibinfo {title} {An anomalous
  topological phase transition in spatial random graphs},\ }\href
  {https://doi.org/10.1038/s42005-022-01023-w} {\bibfield  {journal} {\bibinfo
  {journal} {Communications Physics}\ }\textbf {\bibinfo {volume} {5}},\
  \bibinfo {pages} {245} (\bibinfo {year} {2022})}\BibitemShut {NoStop}%
\bibitem [{\citenamefont {Garc{\'\i}a-P{\'e}rez}\ \emph
  {et~al.}(2019)\citenamefont {Garc{\'\i}a-P{\'e}rez}, \citenamefont {Allard},
  \citenamefont {Serrano},\ and\ \citenamefont
  {Bogu{\~n}{\'a}}}]{GarciaPerez2019}%
  \BibitemOpen
  \bibfield  {author} {\bibinfo {author} {\bibfnamefont {G.}~\bibnamefont
  {Garc{\'\i}a-P{\'e}rez}}, \bibinfo {author} {\bibfnamefont {A.}~\bibnamefont
  {Allard}}, \bibinfo {author} {\bibfnamefont {M.~{\'A}.}\ \bibnamefont
  {Serrano}},\ and\ \bibinfo {author} {\bibfnamefont {M.}~\bibnamefont
  {Bogu{\~n}{\'a}}},\ }\bibfield  {title} {\bibinfo {title} {Mercator:
  uncovering faithful hyperbolic embeddings of complex networks},\ }\href
  {https://doi.org/10.1088/1367-2630/ab57d2} {\bibfield  {journal} {\bibinfo
  {journal} {New Journal of Physics}\ }\textbf {\bibinfo {volume} {21}},\
  \bibinfo {pages} {123033} (\bibinfo {year} {2019})}\BibitemShut {NoStop}%
\bibitem [{\citenamefont {Papadopoulos}\ \emph
  {et~al.}(2015{\natexlab{a}})\citenamefont {Papadopoulos}, \citenamefont
  {Psomas},\ and\ \citenamefont {Krioukov}}]{Papadopoulos2015}%
  \BibitemOpen
  \bibfield  {author} {\bibinfo {author} {\bibfnamefont {F.}~\bibnamefont
  {Papadopoulos}}, \bibinfo {author} {\bibfnamefont {C.}~\bibnamefont
  {Psomas}},\ and\ \bibinfo {author} {\bibfnamefont {D.}~\bibnamefont
  {Krioukov}},\ }\bibfield  {title} {\bibinfo {title} {Network mapping by
  replaying hyperbolic growth},\ }\href
  {https://doi.org/10.1109/TNET.2013.2294052} {\bibfield  {journal} {\bibinfo
  {journal} {IEEE/ACM Transactions on Networking}\ }\textbf {\bibinfo {volume}
  {23}},\ \bibinfo {pages} {198} (\bibinfo {year}
  {2015}{\natexlab{a}})}\BibitemShut {NoStop}%
\bibitem [{\citenamefont {Blasius}\ \emph {et~al.}(2018)\citenamefont
  {Blasius}, \citenamefont {Friedrich}, \citenamefont {Krohmer},\ and\
  \citenamefont {Laue}}]{Blasius2018}%
  \BibitemOpen
  \bibfield  {author} {\bibinfo {author} {\bibfnamefont {T.}~\bibnamefont
  {Blasius}}, \bibinfo {author} {\bibfnamefont {T.}~\bibnamefont {Friedrich}},
  \bibinfo {author} {\bibfnamefont {A.}~\bibnamefont {Krohmer}},\ and\ \bibinfo
  {author} {\bibfnamefont {S.}~\bibnamefont {Laue}},\ }\bibfield  {title}
  {\bibinfo {title} {Efficient embedding of scale-free graphs in the hyperbolic
  plane},\ }\href {https://doi.org/10.1109/TNET.2018.2810186} {\bibfield
  {journal} {\bibinfo  {journal} {IEEE/ACM Transactions on Networking}\
  }\textbf {\bibinfo {volume} {26}},\ \bibinfo {pages} {920} (\bibinfo {year}
  {2018})}\BibitemShut {NoStop}%
\bibitem [{\citenamefont {Muscoloni}\ \emph
  {et~al.}(2017{\natexlab{a}})\citenamefont {Muscoloni}, \citenamefont
  {Thomas}, \citenamefont {Ciucci}, \citenamefont {Bianconi},\ and\
  \citenamefont {Cannistraci}}]{Muscoloni2017}%
  \BibitemOpen
  \bibfield  {author} {\bibinfo {author} {\bibfnamefont {A.}~\bibnamefont
  {Muscoloni}}, \bibinfo {author} {\bibfnamefont {J.~M.}\ \bibnamefont
  {Thomas}}, \bibinfo {author} {\bibfnamefont {S.}~\bibnamefont {Ciucci}},
  \bibinfo {author} {\bibfnamefont {G.}~\bibnamefont {Bianconi}},\ and\
  \bibinfo {author} {\bibfnamefont {C.~V.}\ \bibnamefont {Cannistraci}},\
  }\bibfield  {title} {\bibinfo {title} {Machine learning meets complex
  networks via coalescent embedding in the hyperbolic space},\ }\href
  {https://doi.org/10.1038/s41467-017-01825-5} {\bibfield  {journal} {\bibinfo
  {journal} {Nature Communications}\ }\textbf {\bibinfo {volume} {8}},\
  \bibinfo {pages} {1615} (\bibinfo {year} {2017}{\natexlab{a}})}\BibitemShut
  {NoStop}%
\bibitem [{\citenamefont {van~der Hoorn}\ \emph {et~al.}(2018)\citenamefont
  {van~der Hoorn}, \citenamefont {Lippner},\ and\ \citenamefont
  {Krioukov}}]{vanderHoorn2018}%
  \BibitemOpen
  \bibfield  {author} {\bibinfo {author} {\bibfnamefont {P.}~\bibnamefont
  {van~der Hoorn}}, \bibinfo {author} {\bibfnamefont {G.}~\bibnamefont
  {Lippner}},\ and\ \bibinfo {author} {\bibfnamefont {D.}~\bibnamefont
  {Krioukov}},\ }\bibfield  {title} {\bibinfo {title} {Sparse maximum-entropy
  random graphs with a given power-law degree distribution},\ }\href
  {https://doi.org/10.1007/s10955-017-1887-7} {\bibfield  {journal} {\bibinfo
  {journal} {Journal of Statistical Physics}\ }\textbf {\bibinfo {volume}
  {173}},\ \bibinfo {pages} {806} (\bibinfo {year} {2018})}\BibitemShut
  {NoStop}%
\bibitem [{\citenamefont {Guimerà}\ \emph {et~al.}(2004)\citenamefont
  {Guimerà}, \citenamefont {Sales-Pardo},\ and\ \citenamefont
  {Amaral}}]{Guimera2004}%
  \BibitemOpen
  \bibfield  {author} {\bibinfo {author} {\bibfnamefont {R.}~\bibnamefont
  {Guimerà}}, \bibinfo {author} {\bibfnamefont {M.}~\bibnamefont
  {Sales-Pardo}},\ and\ \bibinfo {author} {\bibfnamefont {L.~A.~N.}\
  \bibnamefont {Amaral}},\ }\bibfield  {title} {\bibinfo {title} {Modularity
  from fluctuations in random graphs and complex networks},\ }\href
  {https://doi.org/10.1103/PhysRevE.70.025101} {\bibfield  {journal} {\bibinfo
  {journal} {Physical Review E}\ }\textbf {\bibinfo {volume} {70}},\ \bibinfo
  {pages} {025101} (\bibinfo {year} {2004})}\BibitemShut {NoStop}%
\bibitem [{\citenamefont {Peixoto}(2022)}]{Peixoto2022}%
  \BibitemOpen
  \bibfield  {author} {\bibinfo {author} {\bibfnamefont {T.~P.}\ \bibnamefont
  {Peixoto}},\ }\bibfield  {title} {\bibinfo {title} {Disentangling homophily,
  community structure, and triadic closure in networks},\ }\href
  {https://doi.org/10.1103/PhysRevX.12.011004} {\bibfield  {journal} {\bibinfo
  {journal} {Physical Review X}\ }\textbf {\bibinfo {volume} {12}},\ \bibinfo
  {pages} {011004} (\bibinfo {year} {2022})}\BibitemShut {NoStop}%
\bibitem [{\citenamefont {Garc{\'\i}a-P{\'e}rez}\ \emph
  {et~al.}(2018{\natexlab{a}})\citenamefont {Garc{\'\i}a-P{\'e}rez},
  \citenamefont {Serrano},\ and\ \citenamefont
  {Bogu{\~n}{\'a}}}]{GarciaPerez2018b}%
  \BibitemOpen
  \bibfield  {author} {\bibinfo {author} {\bibfnamefont {G.}~\bibnamefont
  {Garc{\'\i}a-P{\'e}rez}}, \bibinfo {author} {\bibfnamefont {M.~{\'A}.}\
  \bibnamefont {Serrano}},\ and\ \bibinfo {author} {\bibfnamefont
  {M.}~\bibnamefont {Bogu{\~n}{\'a}}},\ }\bibfield  {title} {\bibinfo {title}
  {Soft communities in similarity space},\ }\href
  {https://doi.org/10.1007/s10955-018-2084-z} {\bibfield  {journal} {\bibinfo
  {journal} {Journal of Statistical Physics}\ }\textbf {\bibinfo {volume}
  {173}},\ \bibinfo {pages} {775} (\bibinfo {year}
  {2018}{\natexlab{a}})}\BibitemShut {NoStop}%
\bibitem [{\citenamefont {Krioukov}\ \emph {et~al.}(2009)\citenamefont
  {Krioukov}, \citenamefont {Papadopoulos}, \citenamefont {Vahdat},\ and\
  \citenamefont {Bogu{\~n}{\'a}}}]{Krioukov2009}%
  \BibitemOpen
  \bibfield  {author} {\bibinfo {author} {\bibfnamefont {D.}~\bibnamefont
  {Krioukov}}, \bibinfo {author} {\bibfnamefont {F.}~\bibnamefont
  {Papadopoulos}}, \bibinfo {author} {\bibfnamefont {A.}~\bibnamefont
  {Vahdat}},\ and\ \bibinfo {author} {\bibfnamefont {M.}~\bibnamefont
  {Bogu{\~n}{\'a}}},\ }\bibfield  {title} {\bibinfo {title} {Curvature and
  temperature of complex networks},\ }\href
  {https://doi.org/10.1103/PhysRevE.80.035101} {\bibfield  {journal} {\bibinfo
  {journal} {Physical Review E}\ }\textbf {\bibinfo {volume} {80}},\ \bibinfo
  {pages} {035101} (\bibinfo {year} {2009})}\BibitemShut {NoStop}%
\bibitem [{\citenamefont {Serrano}\ and\ \citenamefont
  {Bogu{\~n}{\'a}}(2022)}]{Serrano2022}%
  \BibitemOpen
  \bibfield  {author} {\bibinfo {author} {\bibfnamefont {M.~{\'A}.}\
  \bibnamefont {Serrano}}\ and\ \bibinfo {author} {\bibfnamefont
  {M.}~\bibnamefont {Bogu{\~n}{\'a}}},\ }\href
  {https://doi.org/10.1017/9781108865791} {\emph {\bibinfo {title} {The
  Shortest Path to Network Geometry}}}\ (\bibinfo  {publisher} {Cambridge
  University Press},\ \bibinfo {year} {2022})\BibitemShut {NoStop}%
\bibitem [{\citenamefont {Bogu{\~n}{\'a}}\ \emph {et~al.}(2010)\citenamefont
  {Bogu{\~n}{\'a}}, \citenamefont {Papadopoulos},\ and\ \citenamefont
  {Krioukov}}]{Boguna2010}%
  \BibitemOpen
  \bibfield  {author} {\bibinfo {author} {\bibfnamefont {M.}~\bibnamefont
  {Bogu{\~n}{\'a}}}, \bibinfo {author} {\bibfnamefont {F.}~\bibnamefont
  {Papadopoulos}},\ and\ \bibinfo {author} {\bibfnamefont {D.}~\bibnamefont
  {Krioukov}},\ }\bibfield  {title} {\bibinfo {title} {Sustaining the internet
  with hyperbolic mapping},\ }\href {https://doi.org/10.1038/ncomms1063}
  {\bibfield  {journal} {\bibinfo  {journal} {Nature Communications}\ }\textbf
  {\bibinfo {volume} {1}},\ \bibinfo {pages} {62} (\bibinfo {year}
  {2010})}\BibitemShut {NoStop}%
\bibitem [{\citenamefont {Garc{\'\i}a-P{\'e}rez}\ \emph
  {et~al.}(2018{\natexlab{b}})\citenamefont {Garc{\'\i}a-P{\'e}rez},
  \citenamefont {Bogu{\~n}{\'a}},\ and\ \citenamefont
  {Serrano}}]{GarciaPerez2018a}%
  \BibitemOpen
  \bibfield  {author} {\bibinfo {author} {\bibfnamefont {G.}~\bibnamefont
  {Garc{\'\i}a-P{\'e}rez}}, \bibinfo {author} {\bibfnamefont {M.}~\bibnamefont
  {Bogu{\~n}{\'a}}},\ and\ \bibinfo {author} {\bibfnamefont {M.~{\'A}.}\
  \bibnamefont {Serrano}},\ }\bibfield  {title} {\bibinfo {title} {Multiscale
  unfolding of real networks by geometric renormalization},\ }\href
  {https://doi.org/10.1038/s41567-018-0072-5} {\bibfield  {journal} {\bibinfo
  {journal} {Nature Physics}\ }\textbf {\bibinfo {volume} {14}},\ \bibinfo
  {pages} {583} (\bibinfo {year} {2018}{\natexlab{b}})}\BibitemShut {NoStop}%
\bibitem [{\citenamefont {Papadopoulos}\ \emph
  {et~al.}(2015{\natexlab{b}})\citenamefont {Papadopoulos}, \citenamefont
  {Aldecoa},\ and\ \citenamefont {Krioukov}}]{Papadopoulos:2015ub}%
  \BibitemOpen
  \bibfield  {author} {\bibinfo {author} {\bibfnamefont {F.}~\bibnamefont
  {Papadopoulos}}, \bibinfo {author} {\bibfnamefont {R.}~\bibnamefont
  {Aldecoa}},\ and\ \bibinfo {author} {\bibfnamefont {D.}~\bibnamefont
  {Krioukov}},\ }\bibfield  {title} {\bibinfo {title} {Network geometry
  inference using common neighbors},\ }\href
  {https://doi.org/10.1103/PhysRevE.92.022807} {\bibfield  {journal} {\bibinfo
  {journal} {Phys. Rev. E}\ }\textbf {\bibinfo {volume} {92}},\ \bibinfo
  {pages} {022807} (\bibinfo {year} {2015}{\natexlab{b}})}\BibitemShut
  {NoStop}%
\bibitem [{\citenamefont {Bl{\"a}sius}\ \emph {et~al.}(2018)\citenamefont
  {Bl{\"a}sius}, \citenamefont {Friedrich}, \citenamefont {Krohmer},\ and\
  \citenamefont {Laue}}]{Blasius:2018dx}%
  \BibitemOpen
  \bibfield  {author} {\bibinfo {author} {\bibfnamefont {T.}~\bibnamefont
  {Bl{\"a}sius}}, \bibinfo {author} {\bibfnamefont {T.}~\bibnamefont
  {Friedrich}}, \bibinfo {author} {\bibfnamefont {A.}~\bibnamefont {Krohmer}},\
  and\ \bibinfo {author} {\bibfnamefont {S.}~\bibnamefont {Laue}},\ }\bibfield
  {title} {\bibinfo {title} {Efficient embedding of scale-free graphs in the
  hyperbolic plane},\ }\href {https://doi.org/10.1109/TNET.2018.2810186}
  {\bibfield  {journal} {\bibinfo  {journal} {IEEE/ACM Transactions on
  Networking}\ }\textbf {\bibinfo {volume} {26}},\ \bibinfo {pages} {920}
  (\bibinfo {year} {2018})}\BibitemShut {NoStop}%
\bibitem [{\citenamefont {Bl\"{a}sius}\ \emph {et~al.}(2020)\citenamefont
  {Bl\"{a}sius}, \citenamefont {Friedrich}, \citenamefont {Katzmann},\ and\
  \citenamefont {Krohmer}}]{Blasius:2020ko}%
  \BibitemOpen
  \bibfield  {author} {\bibinfo {author} {\bibfnamefont {T.}~\bibnamefont
  {Bl\"{a}sius}}, \bibinfo {author} {\bibfnamefont {T.}~\bibnamefont
  {Friedrich}}, \bibinfo {author} {\bibfnamefont {M.}~\bibnamefont
  {Katzmann}},\ and\ \bibinfo {author} {\bibfnamefont {A.}~\bibnamefont
  {Krohmer}},\ }\bibfield  {title} {\bibinfo {title} {Hyperbolic embeddings for
  near-optimal greedy routing},\ }\bibfield  {journal} {\bibinfo  {journal}
  {ACM J. Exp. Algorithmics}\ }\textbf {\bibinfo {volume} {25}},\ \href
  {https://doi.org/10.1145/3381751} {10.1145/3381751} (\bibinfo {year}
  {2020})\BibitemShut {NoStop}%
\bibitem [{\citenamefont {Kov{\'a}cs}\ and\ \citenamefont
  {Palla}(2021)}]{Kovacs:2021zy}%
  \BibitemOpen
  \bibfield  {author} {\bibinfo {author} {\bibfnamefont {B.}~\bibnamefont
  {Kov{\'a}cs}}\ and\ \bibinfo {author} {\bibfnamefont {G.}~\bibnamefont
  {Palla}},\ }\bibfield  {title} {\bibinfo {title} {Optimisation of the
  coalescent hyperbolic embedding of complex networks},\ }\href
  {https://doi.org/10.1038/s41598-021-87333-5} {\bibfield  {journal} {\bibinfo
  {journal} {Sci. Rep.}\ }\textbf {\bibinfo {volume} {11}},\ \bibinfo {pages}
  {8350} (\bibinfo {year} {2021})}\BibitemShut {NoStop}%
\bibitem [{\citenamefont {Goyal}\ and\ \citenamefont
  {Ferrara}(2018)}]{goyal2018graph}%
  \BibitemOpen
  \bibfield  {author} {\bibinfo {author} {\bibfnamefont {P.}~\bibnamefont
  {Goyal}}\ and\ \bibinfo {author} {\bibfnamefont {E.}~\bibnamefont
  {Ferrara}},\ }\bibfield  {title} {\bibinfo {title} {Graph embedding
  techniques, applications, and performance: A survey},\ }\href@noop {}
  {\bibfield  {journal} {\bibinfo  {journal} {Knowledge-Based Systems}\
  }\textbf {\bibinfo {volume} {151}},\ \bibinfo {pages} {78} (\bibinfo {year}
  {2018})}\BibitemShut {NoStop}%
\bibitem [{\citenamefont {Alanis-Lobato}\ \emph {et~al.}(2016)\citenamefont
  {Alanis-Lobato}, \citenamefont {Mier},\ and\ \citenamefont
  {Andrade-Navarro}}]{AlanisLobato2016}%
  \BibitemOpen
  \bibfield  {author} {\bibinfo {author} {\bibfnamefont {G.}~\bibnamefont
  {Alanis-Lobato}}, \bibinfo {author} {\bibfnamefont {P.}~\bibnamefont
  {Mier}},\ and\ \bibinfo {author} {\bibfnamefont {M.~A.}\ \bibnamefont
  {Andrade-Navarro}},\ }\bibfield  {title} {\bibinfo {title} {Efficient
  embedding of complex networks to hyperbolic space via their laplacian},\
  }\href {https://doi.org/10.1038/srep30108} {\bibfield  {journal} {\bibinfo
  {journal} {Scientific Reports}\ }\textbf {\bibinfo {volume} {6}},\ \bibinfo
  {pages} {30108} (\bibinfo {year} {2016})}\BibitemShut {NoStop}%
\bibitem [{\citenamefont {Muscoloni}\ \emph
  {et~al.}(2017{\natexlab{b}})\citenamefont {Muscoloni}, \citenamefont
  {Thomas}, \citenamefont {Ciucci}, \citenamefont {Bianconi},\ and\
  \citenamefont {Cannistraci}}]{muscoloni2017machine}%
  \BibitemOpen
  \bibfield  {author} {\bibinfo {author} {\bibfnamefont {A.}~\bibnamefont
  {Muscoloni}}, \bibinfo {author} {\bibfnamefont {J.~M.}\ \bibnamefont
  {Thomas}}, \bibinfo {author} {\bibfnamefont {S.}~\bibnamefont {Ciucci}},
  \bibinfo {author} {\bibfnamefont {G.}~\bibnamefont {Bianconi}},\ and\
  \bibinfo {author} {\bibfnamefont {C.~V.}\ \bibnamefont {Cannistraci}},\
  }\bibfield  {title} {\bibinfo {title} {{Machine learning meets complex
  networks via coalescent embedding in the hyperbolic space}},\ }\href
  {https://doi.org/10.1038/s41467-017-01825-5} {\bibfield  {journal} {\bibinfo
  {journal} {Nat. Commun.}\ }\textbf {\bibinfo {volume} {8}},\ \bibinfo {pages}
  {1615} (\bibinfo {year} {2017}{\natexlab{b}})}\BibitemShut {NoStop}%
\bibitem [{\citenamefont {{Keller-Ressel}}\ and\ \citenamefont
  {Nargang}(2020)}]{Keller-Ressel:2020ni}%
  \BibitemOpen
  \bibfield  {author} {\bibinfo {author} {\bibfnamefont {M.}~\bibnamefont
  {{Keller-Ressel}}}\ and\ \bibinfo {author} {\bibfnamefont {S.}~\bibnamefont
  {Nargang}},\ }\bibfield  {title} {\bibinfo {title} {Hydra: a method for
  strain-minimizing hyperbolic embedding of network- and distance-based data},\
  }\href {https://doi.org/10.1093/comnet/cnaa002} {\bibfield  {journal}
  {\bibinfo  {journal} {J. Complex Netw.}\ }\textbf {\bibinfo {volume} {8}},\
  \bibinfo {pages} {cnaa002} (\bibinfo {year} {2020})}\BibitemShut {NoStop}%
\bibitem [{\citenamefont {Belkin}\ and\ \citenamefont
  {Niyogi}(2008)}]{Belkin2008}%
  \BibitemOpen
  \bibfield  {author} {\bibinfo {author} {\bibfnamefont {M.}~\bibnamefont
  {Belkin}}\ and\ \bibinfo {author} {\bibfnamefont {P.}~\bibnamefont
  {Niyogi}},\ }\bibfield  {title} {\bibinfo {title} {Towards a theoretical
  foundation for laplacian-based manifold methods},\ }\href
  {https://doi.org/10.1016/j.jcss.2007.08.006} {\bibfield  {journal} {\bibinfo
  {journal} {Journal of Computer and System Sciences}\ }\textbf {\bibinfo
  {volume} {74}},\ \bibinfo {pages} {1289} (\bibinfo {year}
  {2008})}\BibitemShut {NoStop}%
\bibitem [{\citenamefont {Dunne}\ \emph {et~al.}(2014)\citenamefont {Dunne},
  \citenamefont {Labandeira},\ and\ \citenamefont {Williams}}]{Dunne2014}%
  \BibitemOpen
  \bibfield  {author} {\bibinfo {author} {\bibfnamefont {J.~A.}\ \bibnamefont
  {Dunne}}, \bibinfo {author} {\bibfnamefont {C.~C.}\ \bibnamefont
  {Labandeira}},\ and\ \bibinfo {author} {\bibfnamefont {R.~J.}\ \bibnamefont
  {Williams}},\ }\bibfield  {title} {\bibinfo {title} {Highly resolved early
  eocene food webs show development of modern trophic structure after the
  end-cretaceous extinction},\ }\href {https://doi.org/10.1098/rspb.2013.3280}
  {\bibfield  {journal} {\bibinfo  {journal} {Proceedings of the Royal Society
  B: Biological Sciences}\ }\textbf {\bibinfo {volume} {281}},\ \bibinfo
  {pages} {20133280} (\bibinfo {year} {2014})}\BibitemShut {NoStop}%
\bibitem [{\citenamefont {Cobb}\ and\ \citenamefont {Chen}(2003)}]{Cobb2003}%
  \BibitemOpen
  \bibfield  {author} {\bibinfo {author} {\bibfnamefont {G.~W.}\ \bibnamefont
  {Cobb}}\ and\ \bibinfo {author} {\bibfnamefont {Y.-P.}\ \bibnamefont
  {Chen}},\ }\bibfield  {title} {\bibinfo {title} {An application of markov
  chain monte carlo to community ecology},\ }\href
  {https://doi.org/10.2307/3647877} {\bibfield  {journal} {\bibinfo  {journal}
  {The American Mathematical Monthly}\ }\textbf {\bibinfo {volume} {110}},\
  \bibinfo {pages} {265} (\bibinfo {year} {2003})}\BibitemShut {NoStop}%
\bibitem [{\citenamefont {de~Simon}\ and\ \citenamefont
  {Bogu{\~n}{\'a}}(2012)}]{ColomerdeSimon2012}%
  \BibitemOpen
  \bibfield  {author} {\bibinfo {author} {\bibfnamefont {P.~C.}\ \bibnamefont
  {de~Simon}}\ and\ \bibinfo {author} {\bibfnamefont {M.}~\bibnamefont
  {Bogu{\~n}{\'a}}},\ }\bibfield  {title} {\bibinfo {title} {Clustering of
  random scale-free networks},\ }\href
  {https://doi.org/10.1103/PhysRevE.86.026120} {\bibfield  {journal} {\bibinfo
  {journal} {Physical Review E}\ }\textbf {\bibinfo {volume} {86}},\ \bibinfo
  {pages} {026120} (\bibinfo {year} {2012})}\BibitemShut {NoStop}%
\bibitem [{\citenamefont {Peixoto}(2023)}]{Peixoto2023}%
  \BibitemOpen
  \bibfield  {author} {\bibinfo {author} {\bibfnamefont {T.~P.}\ \bibnamefont
  {Peixoto}},\ }\href {https://doi.org/10.1017/9781009118897} {\emph {\bibinfo
  {title} {Descriptive vs. Inferential Community Detection in Networks}}}\
  (\bibinfo  {publisher} {Cambridge University Press},\ \bibinfo {year}
  {2023})\BibitemShut {NoStop}%
\bibitem [{\citenamefont {van~der Kolk}\ \emph {et~al.}()\citenamefont {van~der
  Kolk}, \citenamefont {\'{A}ngeles Serrano},\ and\ \citenamefont
  {Bogu{\~{n}}{\'{a}}}}]{supp}%
  \BibitemOpen
  \bibfield  {author} {\bibinfo {author} {\bibfnamefont {J.}~\bibnamefont
  {van~der Kolk}}, \bibinfo {author} {\bibfnamefont {M.}~\bibnamefont
  {\'{A}ngeles Serrano}},\ and\ \bibinfo {author} {\bibfnamefont
  {M.}~\bibnamefont {Bogu{\~{n}}{\'{a}}}},\ }\href@noop {} {\bibinfo {title}
  {Supplementary information for random graphs and real networks with weak
  geometric coupling}}\BibitemShut {NoStop}%
\bibitem [{\citenamefont {Zagar}\ \emph {et~al.}(2011)\citenamefont {Zagar},
  \citenamefont {Mulas}, \citenamefont {Garagna}, \citenamefont {Zuccotti},
  \citenamefont {Bellazzi},\ and\ \citenamefont {Zupan}}]{Zagar2011}%
  \BibitemOpen
  \bibfield  {author} {\bibinfo {author} {\bibfnamefont {L.}~\bibnamefont
  {Zagar}}, \bibinfo {author} {\bibfnamefont {F.}~\bibnamefont {Mulas}},
  \bibinfo {author} {\bibfnamefont {S.}~\bibnamefont {Garagna}}, \bibinfo
  {author} {\bibfnamefont {M.}~\bibnamefont {Zuccotti}}, \bibinfo {author}
  {\bibfnamefont {R.}~\bibnamefont {Bellazzi}},\ and\ \bibinfo {author}
  {\bibfnamefont {B.}~\bibnamefont {Zupan}},\ }\bibfield  {title} {\bibinfo
  {title} {Stage prediction of embryonic stem cell differentiation from
  genome-wide expression data},\ }\href
  {https://doi.org/10.1093/bioinformatics/btr422} {\bibfield  {journal}
  {\bibinfo  {journal} {Bioinformatics}\ }\textbf {\bibinfo {volume} {27}},\
  \bibinfo {pages} {2546} (\bibinfo {year} {2011})}\BibitemShut {NoStop}%
\bibitem [{\citenamefont {Ulanowicz}\ and\ \citenamefont
  {DeAngelis}(2005)}]{Ulanowicz2005}%
  \BibitemOpen
  \bibfield  {author} {\bibinfo {author} {\bibfnamefont {R.~E.}\ \bibnamefont
  {Ulanowicz}}\ and\ \bibinfo {author} {\bibfnamefont {D.~L.}\ \bibnamefont
  {DeAngelis}},\ }\href@noop {} {\bibinfo {title} {Network analysis of trophic
  dynamics in south florida ecosystems}} (\bibinfo {year} {2005})\BibitemShut
  {NoStop}%
\bibitem [{\citenamefont {Milo}\ \emph {et~al.}(2004)\citenamefont {Milo},
  \citenamefont {Itzkovitz}, \citenamefont {Kashtan}, \citenamefont {Levitt},
  \citenamefont {Shen-Orr}, \citenamefont {Ayzenshtat}, \citenamefont
  {Sheffer},\ and\ \citenamefont {Alon}}]{Milo2004}%
  \BibitemOpen
  \bibfield  {author} {\bibinfo {author} {\bibfnamefont {R.}~\bibnamefont
  {Milo}}, \bibinfo {author} {\bibfnamefont {S.}~\bibnamefont {Itzkovitz}},
  \bibinfo {author} {\bibfnamefont {N.}~\bibnamefont {Kashtan}}, \bibinfo
  {author} {\bibfnamefont {R.}~\bibnamefont {Levitt}}, \bibinfo {author}
  {\bibfnamefont {S.}~\bibnamefont {Shen-Orr}}, \bibinfo {author}
  {\bibfnamefont {I.}~\bibnamefont {Ayzenshtat}}, \bibinfo {author}
  {\bibfnamefont {M.}~\bibnamefont {Sheffer}},\ and\ \bibinfo {author}
  {\bibfnamefont {U.}~\bibnamefont {Alon}},\ }\bibfield  {title} {\bibinfo
  {title} {Superfamilies of evolved and designed networks},\ }\href
  {https://doi.org/10.1126/science.1089167} {\bibfield  {journal} {\bibinfo
  {journal} {Science}\ }\textbf {\bibinfo {volume} {303}},\ \bibinfo {pages}
  {1538} (\bibinfo {year} {2004})}\BibitemShut {NoStop}%
\bibitem [{\citenamefont {Huss}\ and\ \citenamefont {Holme}(2007)}]{Huss2007}%
  \BibitemOpen
  \bibfield  {author} {\bibinfo {author} {\bibfnamefont {M.}~\bibnamefont
  {Huss}}\ and\ \bibinfo {author} {\bibfnamefont {P.}~\bibnamefont {Holme}},\
  }\bibfield  {title} {\bibinfo {title} {Currency and commodity metabolites:
  their identification and relation to the modularity of metabolic networks},\
  }\href {https://doi.org/10.1049/iet-syb:20060077} {\bibfield  {journal}
  {\bibinfo  {journal} {IET Systems Biology}\ }\textbf {\bibinfo {volume}
  {1}},\ \bibinfo {pages} {280} (\bibinfo {year} {2007})}\BibitemShut {NoStop}%
\bibitem [{\citenamefont {Kunegis}(2013)}]{Kunegis2013}%
  \BibitemOpen
  \bibfield  {author} {\bibinfo {author} {\bibfnamefont {J.}~\bibnamefont
  {Kunegis}},\ }\bibfield  {title} {\bibinfo {title} {Konect}\ }(\bibinfo
  {publisher} {ACM},\ \bibinfo {year} {2013})\ pp.\ \bibinfo {pages}
  {1343--1350}\BibitemShut {NoStop}%
\bibitem [{\citenamefont {Hu}\ \emph {et~al.}(2018)\citenamefont {Hu},
  \citenamefont {Vinayagam}, \citenamefont {Nand}, \citenamefont {Comjean},
  \citenamefont {Chung}, \citenamefont {Hao}, \citenamefont {Mohr},\ and\
  \citenamefont {Perrimon}}]{Hu2018}%
  \BibitemOpen
  \bibfield  {author} {\bibinfo {author} {\bibfnamefont {Y.}~\bibnamefont
  {Hu}}, \bibinfo {author} {\bibfnamefont {A.}~\bibnamefont {Vinayagam}},
  \bibinfo {author} {\bibfnamefont {A.}~\bibnamefont {Nand}}, \bibinfo {author}
  {\bibfnamefont {A.}~\bibnamefont {Comjean}}, \bibinfo {author} {\bibfnamefont
  {V.}~\bibnamefont {Chung}}, \bibinfo {author} {\bibfnamefont
  {T.}~\bibnamefont {Hao}}, \bibinfo {author} {\bibfnamefont {S.~E.}\
  \bibnamefont {Mohr}},\ and\ \bibinfo {author} {\bibfnamefont
  {N.}~\bibnamefont {Perrimon}},\ }\bibfield  {title} {\bibinfo {title}
  {Molecular interaction search tool (mist): an integrated resource for mining
  gene and protein interaction data},\ }\href
  {https://doi.org/10.1093/nar/gkx1116} {\bibfield  {journal} {\bibinfo
  {journal} {Nucleic Acids Research}\ }\textbf {\bibinfo {volume} {46}},\
  \bibinfo {pages} {D567} (\bibinfo {year} {2018})}\BibitemShut {NoStop}%
\bibitem [{\citenamefont {Domenico}\ \emph {et~al.}(2015)\citenamefont
  {Domenico}, \citenamefont {Porter},\ and\ \citenamefont
  {Arenas}}]{DeDomenico2015}%
  \BibitemOpen
  \bibfield  {author} {\bibinfo {author} {\bibfnamefont {M.~D.}\ \bibnamefont
  {Domenico}}, \bibinfo {author} {\bibfnamefont {M.~A.}\ \bibnamefont
  {Porter}},\ and\ \bibinfo {author} {\bibfnamefont {A.}~\bibnamefont
  {Arenas}},\ }\bibfield  {title} {\bibinfo {title} {Muxviz: a tool for
  multilayer analysis and visualization of networks},\ }\href
  {https://doi.org/10.1093/comnet/cnu038} {\bibfield  {journal} {\bibinfo
  {journal} {Journal of Complex Networks}\ }\textbf {\bibinfo {volume} {3}},\
  \bibinfo {pages} {159} (\bibinfo {year} {2015})}\BibitemShut {NoStop}%
\bibitem [{\citenamefont {Ripeanu}\ and\ \citenamefont
  {Foster}(2002)}]{Ripeanu2002}%
  \BibitemOpen
  \bibfield  {author} {\bibinfo {author} {\bibfnamefont {M.}~\bibnamefont
  {Ripeanu}}\ and\ \bibinfo {author} {\bibfnamefont {I.}~\bibnamefont
  {Foster}},\ }\href {https://doi.org/10.1007/3-540-45748-8_8} {\bibinfo
  {title} {Mapping the gnutella network: Macroscopic properties of large-scale
  peer-to-peer systems}} (\bibinfo {year} {2002})\BibitemShut {NoStop}%
\bibitem [{\citenamefont {Domenico}\ \emph {et~al.}(2014)\citenamefont
  {Domenico}, \citenamefont {Sol{\'e}-Ribalta}, \citenamefont {G{\'o}mez},\
  and\ \citenamefont {Arenas}}]{DeDomenico2014}%
  \BibitemOpen
  \bibfield  {author} {\bibinfo {author} {\bibfnamefont {M.~D.}\ \bibnamefont
  {Domenico}}, \bibinfo {author} {\bibfnamefont {A.}~\bibnamefont
  {Sol{\'e}-Ribalta}}, \bibinfo {author} {\bibfnamefont {S.}~\bibnamefont
  {G{\'o}mez}},\ and\ \bibinfo {author} {\bibfnamefont {A.}~\bibnamefont
  {Arenas}},\ }\bibfield  {title} {\bibinfo {title} {Navigability of
  interconnected networks under random failures},\ }\href
  {https://doi.org/10.1073/pnas.1318469111} {\bibfield  {journal} {\bibinfo
  {journal} {Proceedings of the National Academy of Sciences}\ }\textbf
  {\bibinfo {volume} {111}},\ \bibinfo {pages} {8351} (\bibinfo {year}
  {2014})}\BibitemShut {NoStop}%
\bibitem [{\citenamefont {Knight}\ \emph {et~al.}(2011)\citenamefont {Knight},
  \citenamefont {Nguyen}, \citenamefont {Falkner}, \citenamefont {Bowden},\
  and\ \citenamefont {Roughan}}]{Knight2011}%
  \BibitemOpen
  \bibfield  {author} {\bibinfo {author} {\bibfnamefont {S.}~\bibnamefont
  {Knight}}, \bibinfo {author} {\bibfnamefont {H.~X.}\ \bibnamefont {Nguyen}},
  \bibinfo {author} {\bibfnamefont {N.}~\bibnamefont {Falkner}}, \bibinfo
  {author} {\bibfnamefont {R.}~\bibnamefont {Bowden}},\ and\ \bibinfo {author}
  {\bibfnamefont {M.}~\bibnamefont {Roughan}},\ }\bibfield  {title} {\bibinfo
  {title} {The internet topology zoo},\ }\href
  {https://doi.org/10.1109/JSAC.2011.111002} {\bibfield  {journal} {\bibinfo
  {journal} {IEEE Journal on Selected Areas in Communications}\ }\textbf
  {\bibinfo {volume} {29}},\ \bibinfo {pages} {1765} (\bibinfo {year}
  {2011})}\BibitemShut {NoStop}%
\bibitem [{\citenamefont {Leskovec}\ \emph {et~al.}(2010)\citenamefont
  {Leskovec}, \citenamefont {Huttenlocher},\ and\ \citenamefont
  {Kleinberg}}]{Leskovec2010}%
  \BibitemOpen
  \bibfield  {author} {\bibinfo {author} {\bibfnamefont {J.}~\bibnamefont
  {Leskovec}}, \bibinfo {author} {\bibfnamefont {D.}~\bibnamefont
  {Huttenlocher}},\ and\ \bibinfo {author} {\bibfnamefont {J.}~\bibnamefont
  {Kleinberg}},\ }\bibfield  {title} {\bibinfo {title} {Signed networks in
  social media}\ }(\bibinfo  {publisher} {ACM},\ \bibinfo {year} {2010})\ pp.\
  \bibinfo {pages} {1361--1370}\BibitemShut {NoStop}%
\bibitem [{\citenamefont {Paranjape}\ \emph {et~al.}(2017)\citenamefont
  {Paranjape}, \citenamefont {Benson},\ and\ \citenamefont
  {Leskovec}}]{Paranjape2017}%
  \BibitemOpen
  \bibfield  {author} {\bibinfo {author} {\bibfnamefont {A.}~\bibnamefont
  {Paranjape}}, \bibinfo {author} {\bibfnamefont {A.~R.}\ \bibnamefont
  {Benson}},\ and\ \bibinfo {author} {\bibfnamefont {J.}~\bibnamefont
  {Leskovec}},\ }\bibfield  {title} {\bibinfo {title} {Motifs in temporal
  networks}\ }(\bibinfo  {publisher} {ACM},\ \bibinfo {year} {2017})\ pp.\
  \bibinfo {pages} {601--610}\BibitemShut {NoStop}%
\end{thebibliography}%

\end{document}



\baselineskip24pt


\maketitle 

\tableofcontents
\section{Supplementary Figures}
In the following section we will give the embedding and renormalization results for several real networks. The network properties can be found in Tab.~1 in the main text. Similarly, a brief description of the network can be found in App.~B.
\newpage
\begin{figure}[h]
	\centering
	\includegraphics[width=1\textwidth]{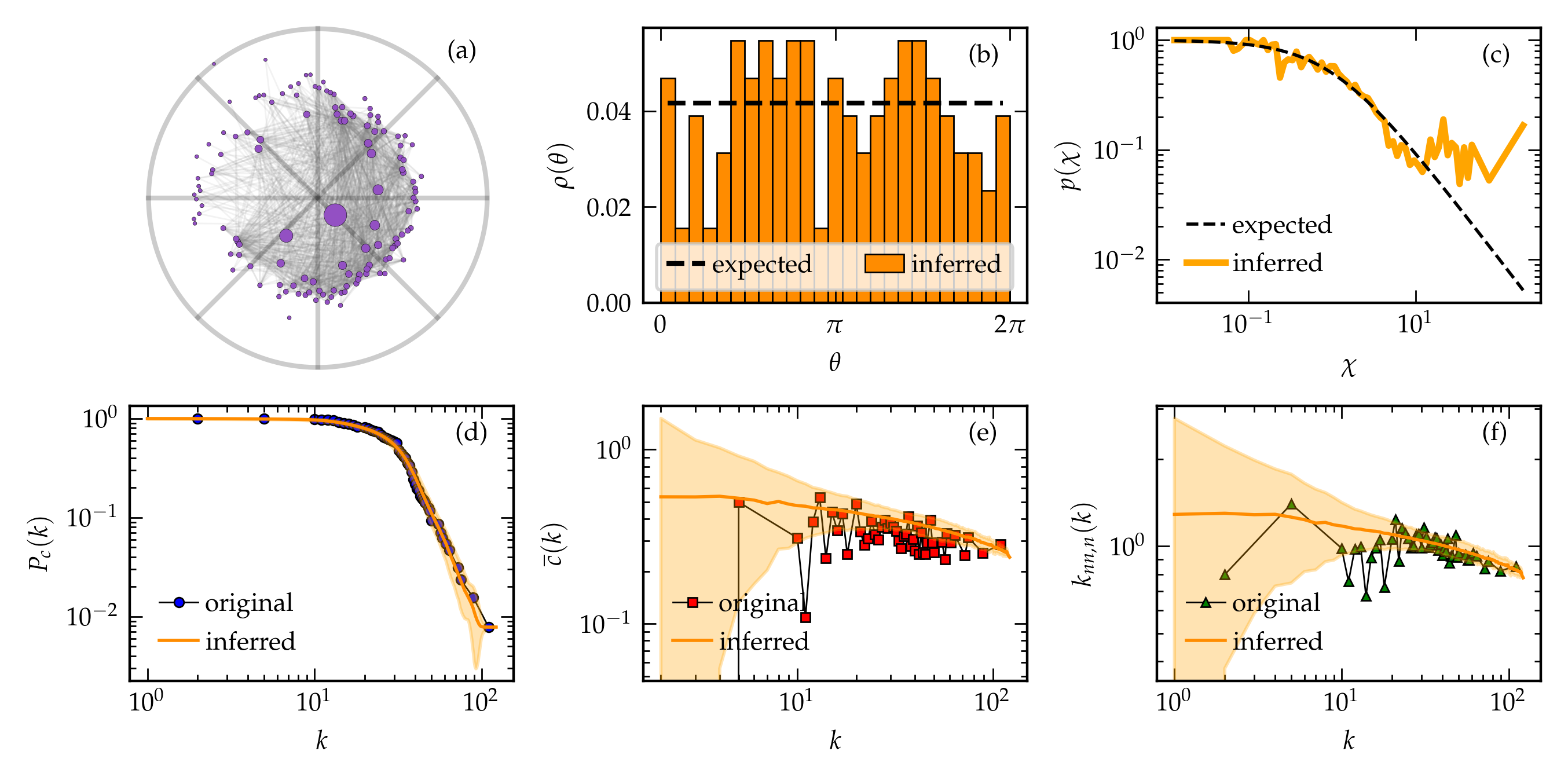}
	\vspace{-8mm}
	\caption{Summary of the results of Mercator for the Foodweb-Eocene network. \textbf{(a)} Representation of the embedding in the hyperbolic plane as defined by the $\mathbb{H}^2$-model. The top 100\% most geometric edges are shown. \textbf{(b)} Comparison between the expected and inferred densities of nodes along the circle. \textbf{(c)} Comparison between the probability distribution as expected based on the model (expected) as well as the actual distribution based on the inferred coordinates (inferred). The reproduction of the topological properties is also given: \textbf{(d)} the complementary cumulative degree distribution, \textbf{(e)} the average local clustering coefficient per degree class and \textbf{(f)} the degree-degree correlations per degree class. The inferred results are obtained by generating 100 realizations of the $\mathbb{S}^1$-model based on the inferred coordinates. The orange shaded regions represent the $2\sigma$ confidence interval. }
	\label{Sfig:fooweb_baywet}
\end{figure}

\newpage


\begin{figure}[h]
	\centering
	\includegraphics[width=1\textwidth]{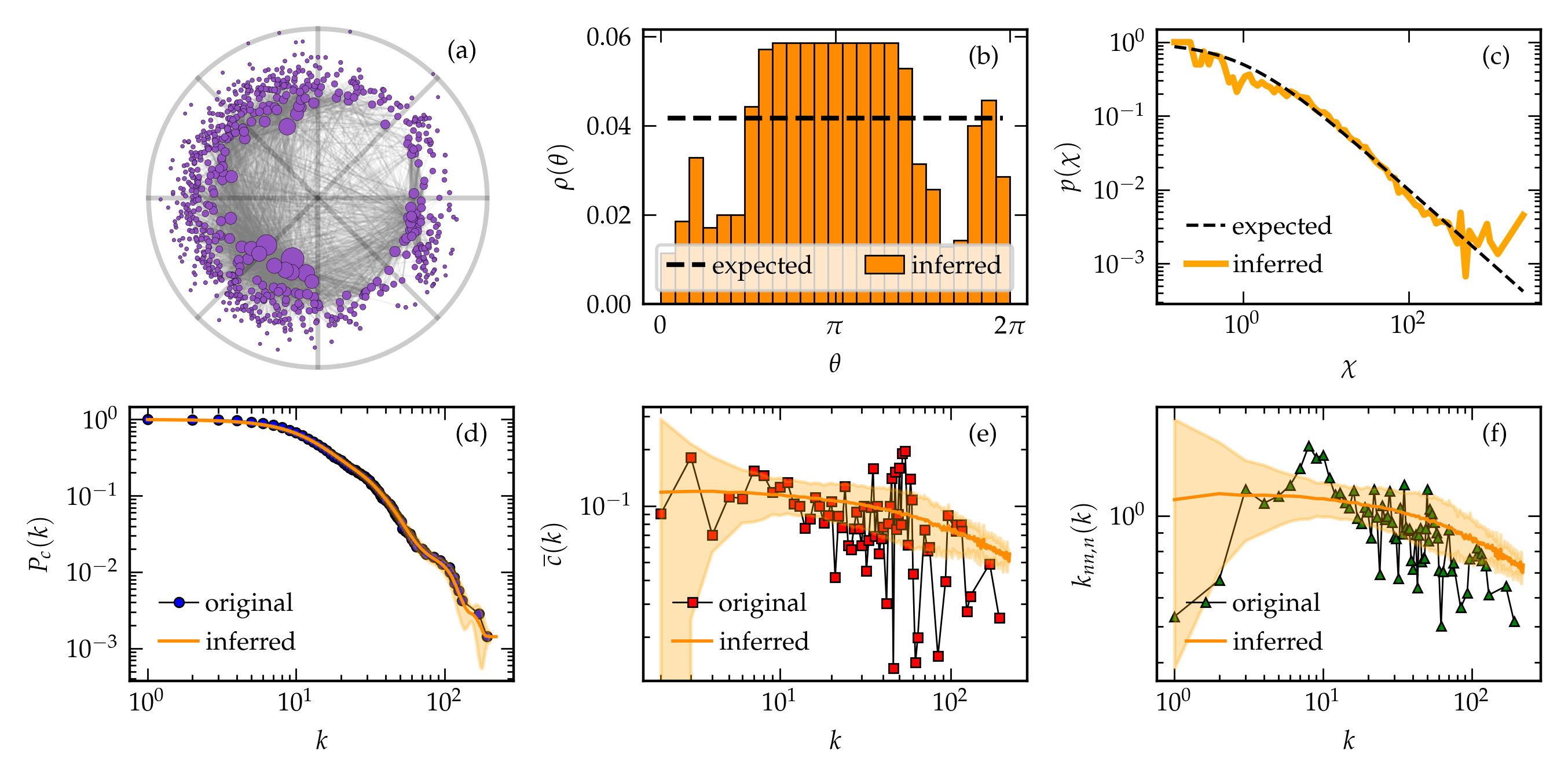}
	\vspace{-8mm}
	\caption{Summary of the results of Mercator for the Foodweb-Wetland network. \textbf{(a)} Representation of the embedding in the hyperbolic plane as defined by the $\mathbb{H}^2$-model. The top 50\% most geometric edges are shown. \textbf{(b)} Comparison between the expected and inferred densities of nodes along the circle. \textbf{(c)} Comparison between the probability distribution as expected based on the model (expected) as well as the actual distribution based on the inferred coordinates (inferred). The reproduction of the topological properties is also given: \textbf{(d)} the complementary cumulative degree distribution, \textbf{(e)} the average local clustering coefficient per degree class and \textbf{(f)} the degree-degree correlations per degree class. The inferred results are obtained by generating 100 realizations of the $\mathbb{S}^1$-model based on the inferred coordinates. The orange shaded regions represent the $2\sigma$ confidence interval. }
	\label{Sfig:fooweb_baywet}
\end{figure}

\clearpage


\begin{figure}[b]
	\centering
	\includegraphics[width=1\textwidth]{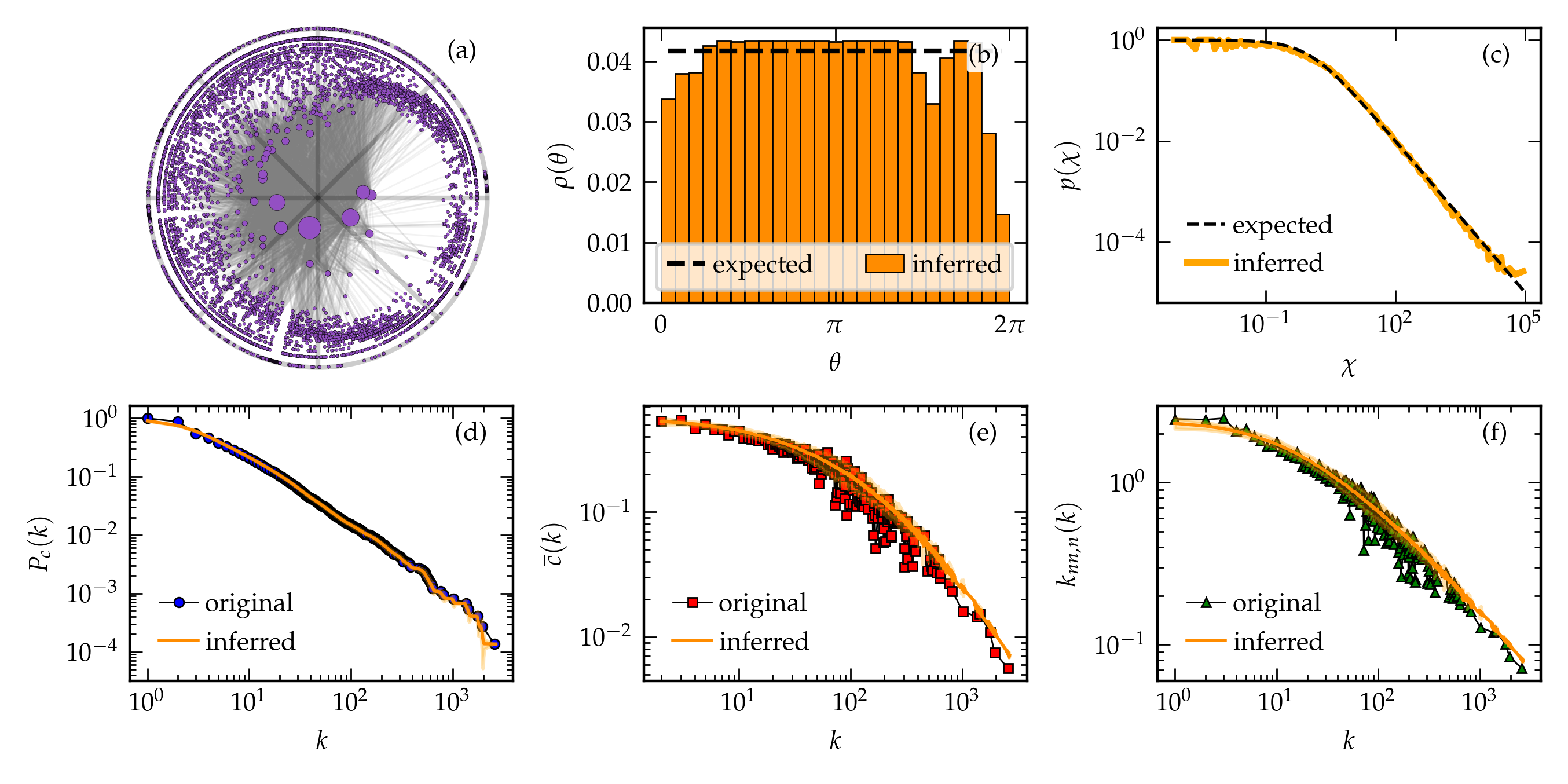}
	\vspace{-8mm}
	\caption{Summary of the results of Mercator for the WordAdjacency-English network. \textbf{(a)} Representation of the embedding in the hyperbolic plane as defined by the $\mathbb{H}^2$-model. The top 10\% most geometric edges are shown. \textbf{(b)} Comparison between the expected and inferred densities of nodes along the circle. \textbf{(c)} Comparison between the probability distribution as expected based on the model (expected) as well as the actual distribution based on the inferred coordinates (inferred). The reproduction of the topological properties is also given: \textbf{(d)} the complementary cumulative degree distribution, \textbf{(e)} the average local clustering coefficient per degree class and \textbf{(f)} the degree-degree correlations per degree class. The inferred results are obtained by generating 100 realizations of the $\mathbb{S}^1$-model based on the inferred coordinates. The orange shaded regions represent the $2\sigma$ confidence interval. }
	\label{Sfig:fooweb_baywet}
\end{figure}

\clearpage


\begin{figure}[b]
	\centering
	\includegraphics[width=1\textwidth]{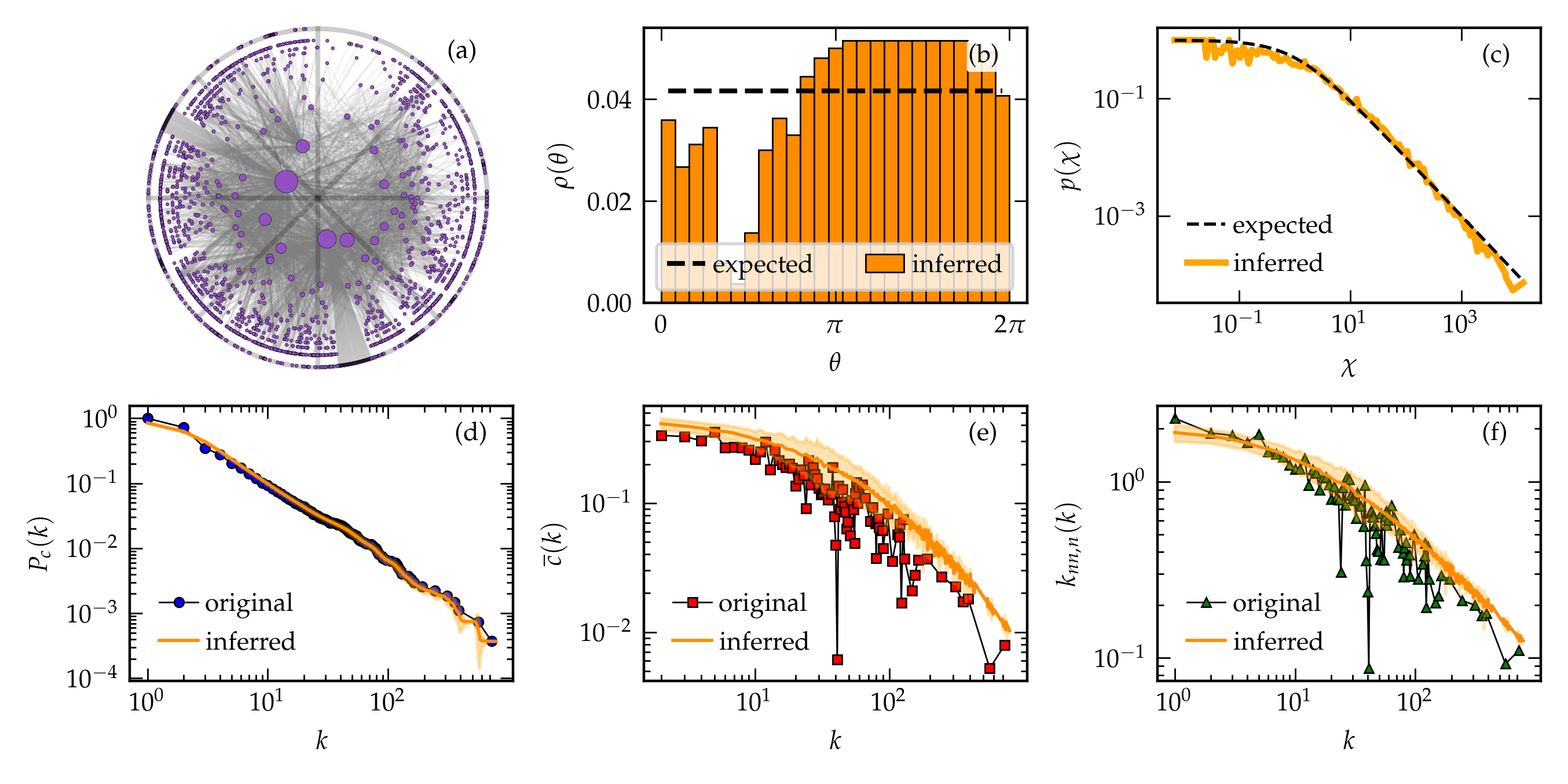}
	\vspace{-8mm}
	\caption{Summary of the results of Mercator for the WordAdjacency–Japanese network. \textbf{(a)} Representation of the embedding in the hyperbolic plane as defined by the $\mathbb{H}^2$-model. The top 50\% most geometric edges are shown. \textbf{(b)} Comparison between the expected and inferred densities of nodes along the circle. \textbf{(c)} Comparison between the probability distribution as expected based on the model (expected) as well as the actual distribution based on the inferred coordinates (inferred). The reproduction of the topological properties is also given: \textbf{(d)} the complementary cumulative degree distribution, \textbf{(e)} the average local clustering coefficient per degree class and \textbf{(f)} the degree-degree correlations per degree class. The inferred results are obtained by generating 100 realizations of the $\mathbb{S}^1$-model based on the inferred coordinates. The orange shaded regions represent the $2\sigma$ confidence interval. }
	\label{Sfig:fooweb_baywet}
\end{figure}


\begin{figure}[b]
	\centering
	\includegraphics[width=1\textwidth]{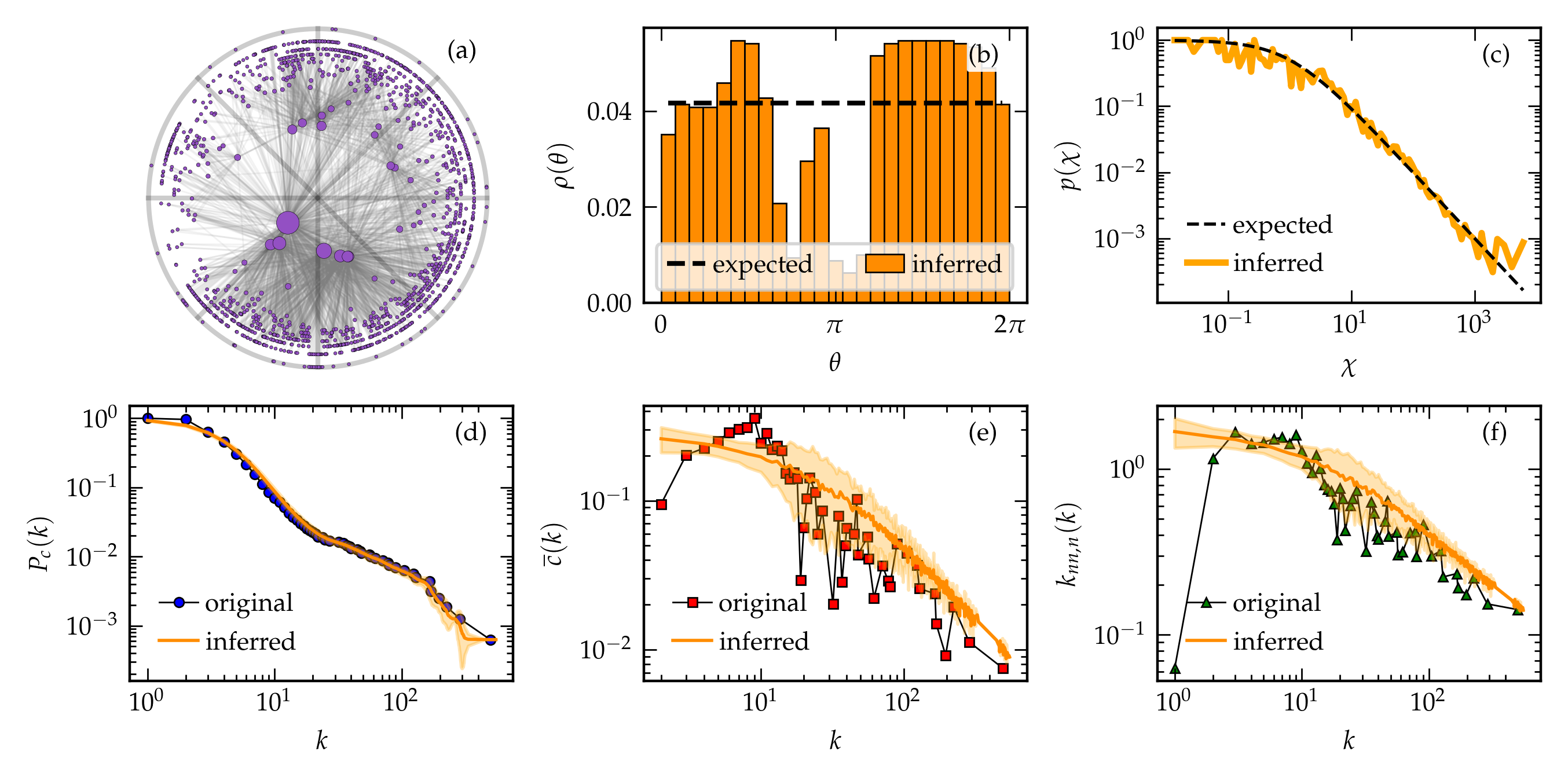}
	\vspace{-8mm}
	\caption{Summary of the results of Mercator for the MB–R.norvegicus network. \textbf{(a)} Representation of the embedding in the hyperbolic plane as defined by the $\mathbb{H}^2$-model. The top 50\% most geometric edges are shown. \textbf{(b)} Comparison between the expected and inferred densities of nodes along the circle. \textbf{(c)} Comparison between the probability distribution as expected based on the model (expected) as well as the actual distribution based on the inferred coordinates (inferred). The reproduction of the topological properties is also given: \textbf{(d)} the complementary cumulative degree distribution, \textbf{(e)} the average local clustering coefficient per degree class and \textbf{(f)} the degree-degree correlations per degree class. The inferred results are obtained by generating 100 realizations of the $\mathbb{S}^1$-model based on the inferred coordinates. The orange shaded regions represent the $2\sigma$ confidence interval. }
	\label{Sfig:fooweb_baywet}
\end{figure}


\begin{figure}[b]
	\centering
	\includegraphics[width=1\textwidth]{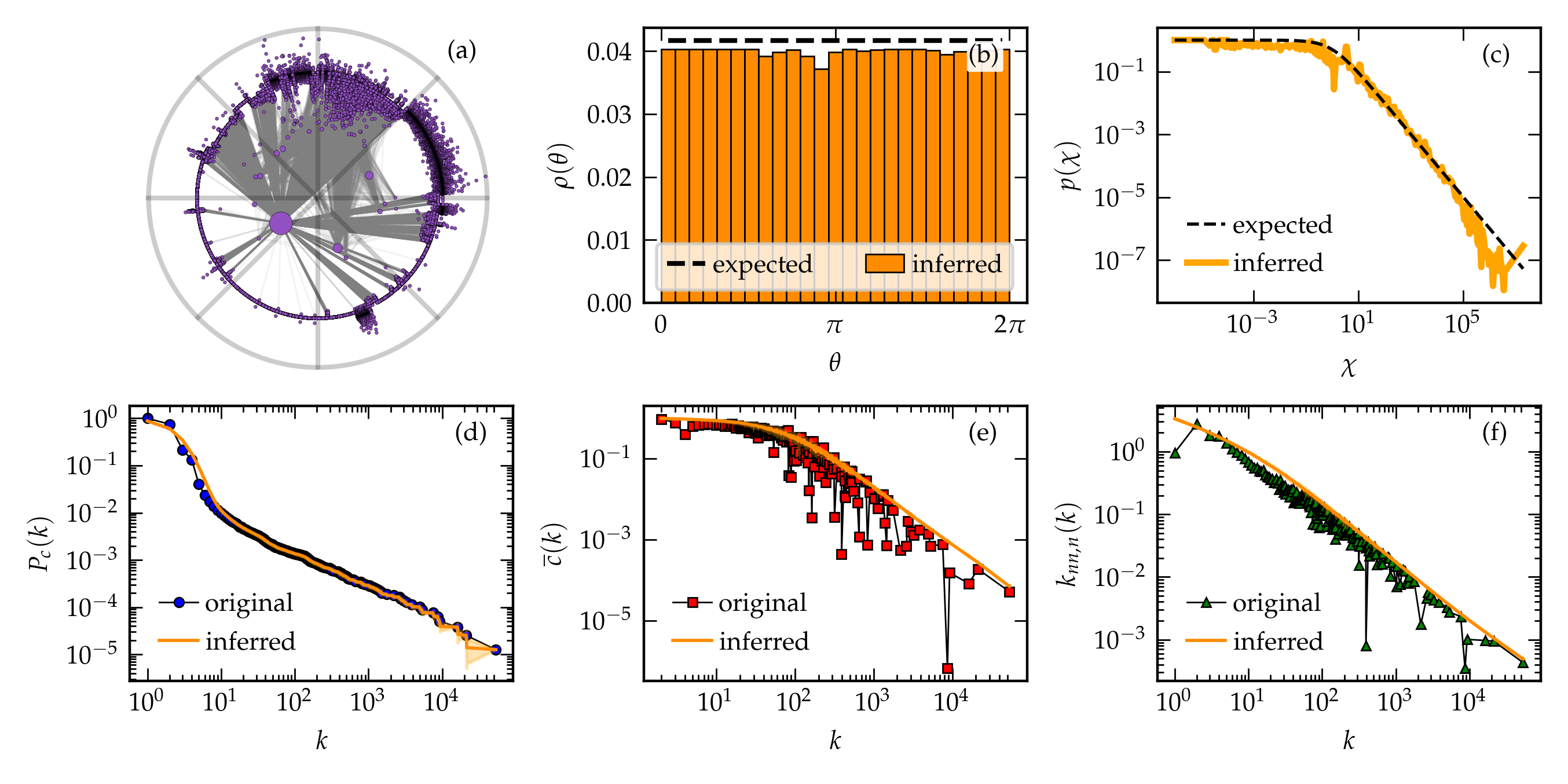}
	\vspace{-8mm}
	\caption{Summary of the results of Mercator for the WikiTalk–Catalan network. \textbf{(a)} Representation of the embedding in the hyperbolic plane as defined by the $\mathbb{H}^2$-model. The top 10\% most geometric edges are shown. \textbf{(b)} Comparison between the expected and inferred densities of nodes along the circle. \textbf{(c)} Comparison between the probability distribution as expected based on the model (expected) as well as the actual distribution based on the inferred coordinates (inferred). The reproduction of the topological properties is also given: \textbf{(d)} the complementary cumulative degree distribution, \textbf{(e)} the average local clustering coefficient per degree class and \textbf{(f)} the degree-degree correlations per degree class. The inferred results are obtained by generating 100 realizations of the $\mathbb{S}^1$-model based on the inferred coordinates. The orange shaded regions represent the $2\sigma$ confidence interval. }
	\label{Sfig:fooweb_baywet}
\end{figure}


\begin{figure}[b]
	\centering
	\includegraphics[width=1\textwidth]{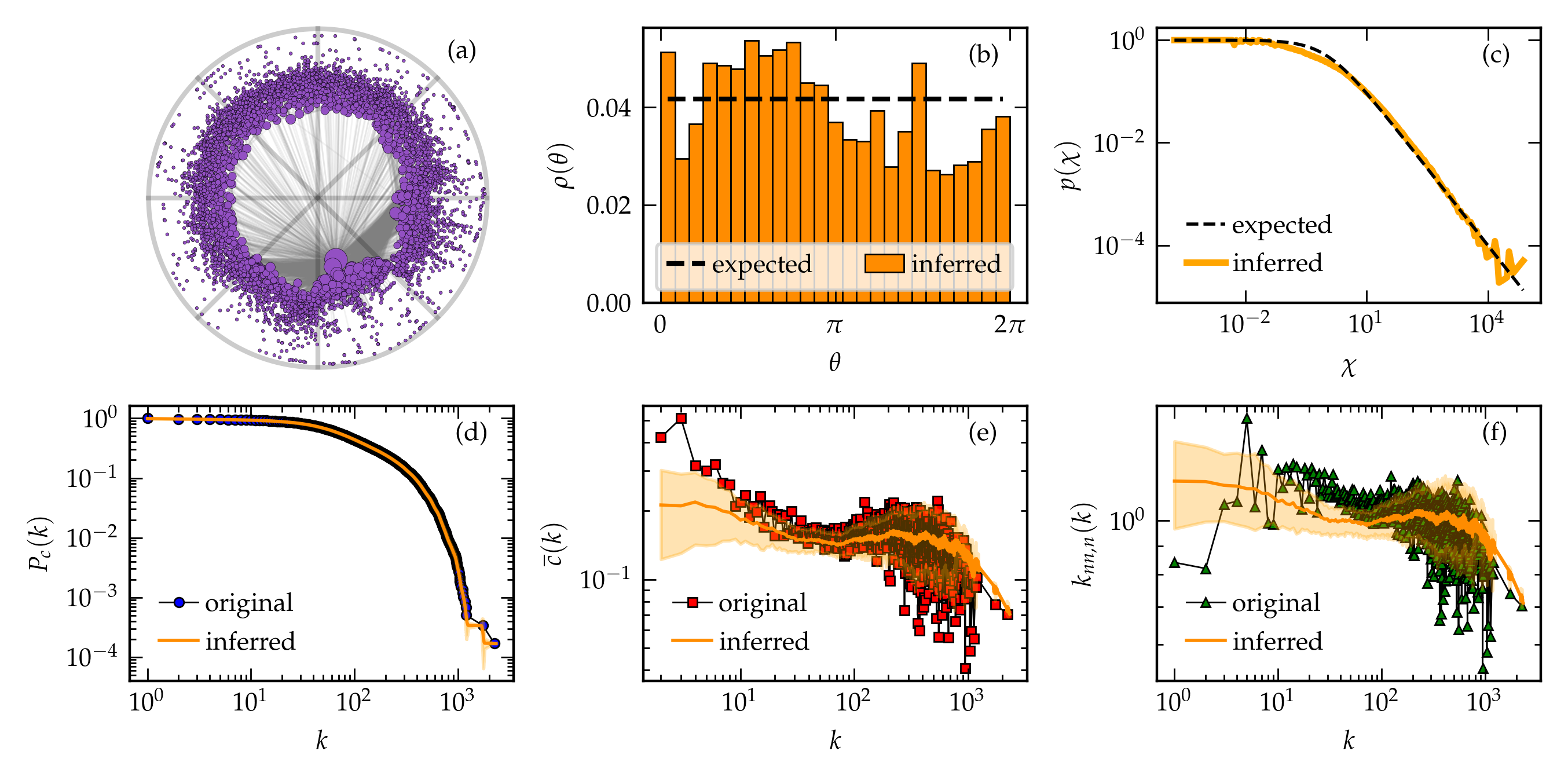}
	\vspace{-8mm}
	\caption{Summary of the results of Mercator for the GI–S.cerevisiae network. \textbf{(a)} Representation of the embedding in the hyperbolic plane as defined by the $\mathbb{H}^2$-model. The top 1\% most geometric edges are shown. \textbf{(b)} Comparison between the expected and inferred densities of nodes along the circle. \textbf{(c)} Comparison between the probability distribution as expected based on the model (expected) as well as the actual distribution based on the inferred coordinates (inferred). The reproduction of the topological properties is also given: \textbf{(d)} the complementary cumulative degree distribution, \textbf{(e)} the average local clustering coefficient per degree class and \textbf{(f)} the degree-degree correlations per degree class. The inferred results are obtained by generating 100 realizations of the $\mathbb{S}^1$-model based on the inferred coordinates. The orange shaded regions represent the $2\sigma$ confidence interval. }
	\label{Sfig:fooweb_baywet}
\end{figure}


\begin{figure}[b]
	\centering
	\includegraphics[width=1\textwidth]{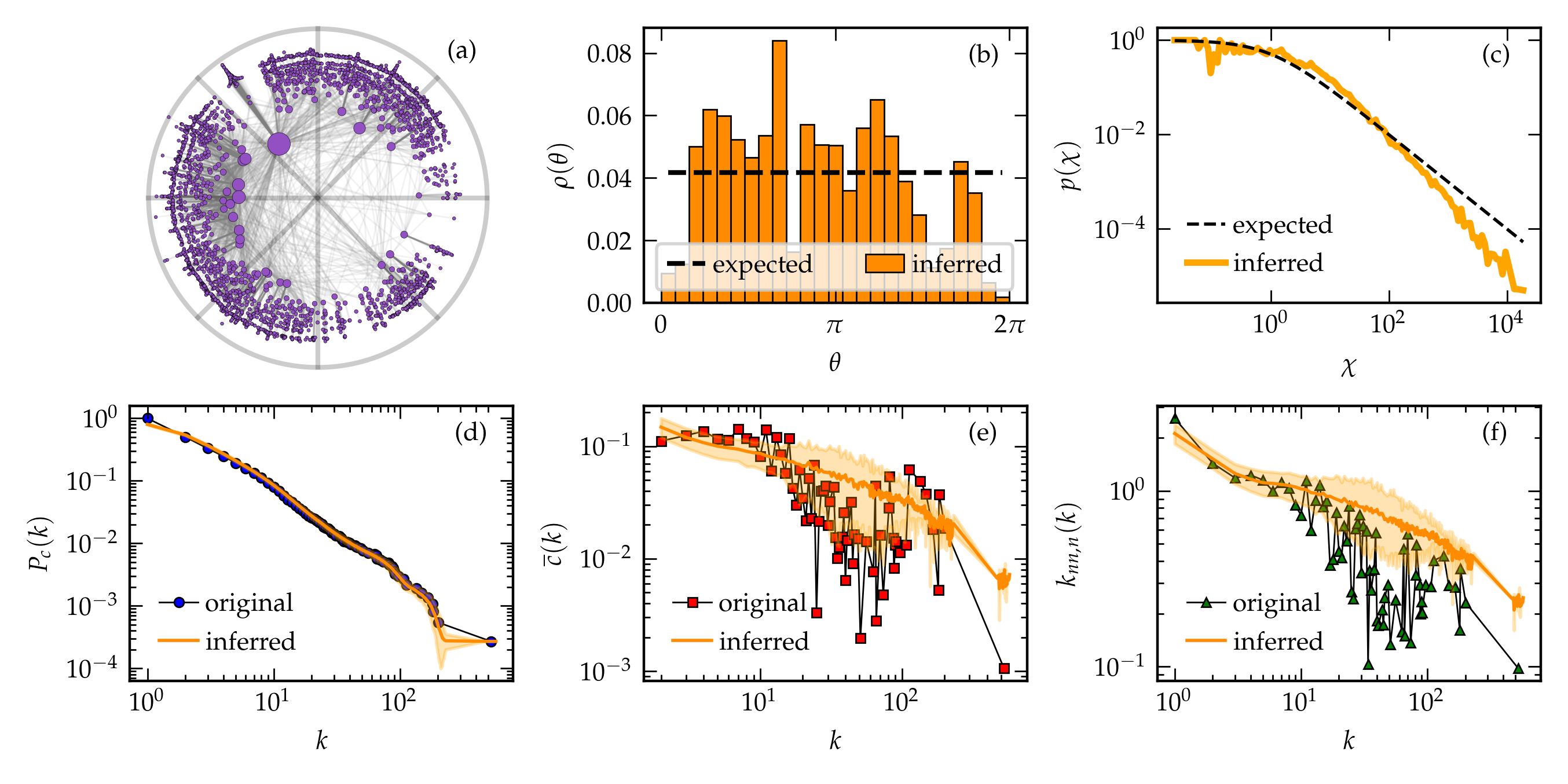}
	\vspace{-8mm}
	\caption{Summary of the results of Mercator for the GMP–C.elegans network. \textbf{(a)} Representation of the embedding in the hyperbolic plane as defined by the $\mathbb{H}^2$-model. The top 50\% most geometric edges are shown. \textbf{(b)} Comparison between the expected and inferred densities of nodes along the circle. \textbf{(c)} Comparison between the probability distribution as expected based on the model (expected) as well as the actual distribution based on the inferred coordinates (inferred). The reproduction of the topological properties is also given: \textbf{(d)} the complementary cumulative degree distribution, \textbf{(e)} the average local clustering coefficient per degree class and \textbf{(f)} the degree-degree correlations per degree class. The inferred results are obtained by generating 100 realizations of the $\mathbb{S}^1$-model based on the inferred coordinates. The orange shaded regions represent the $2\sigma$ confidence interval. }
	\label{Sfig:fooweb_baywet}
\end{figure}


\begin{figure}[b]
	\centering
	\includegraphics[width=1\textwidth]{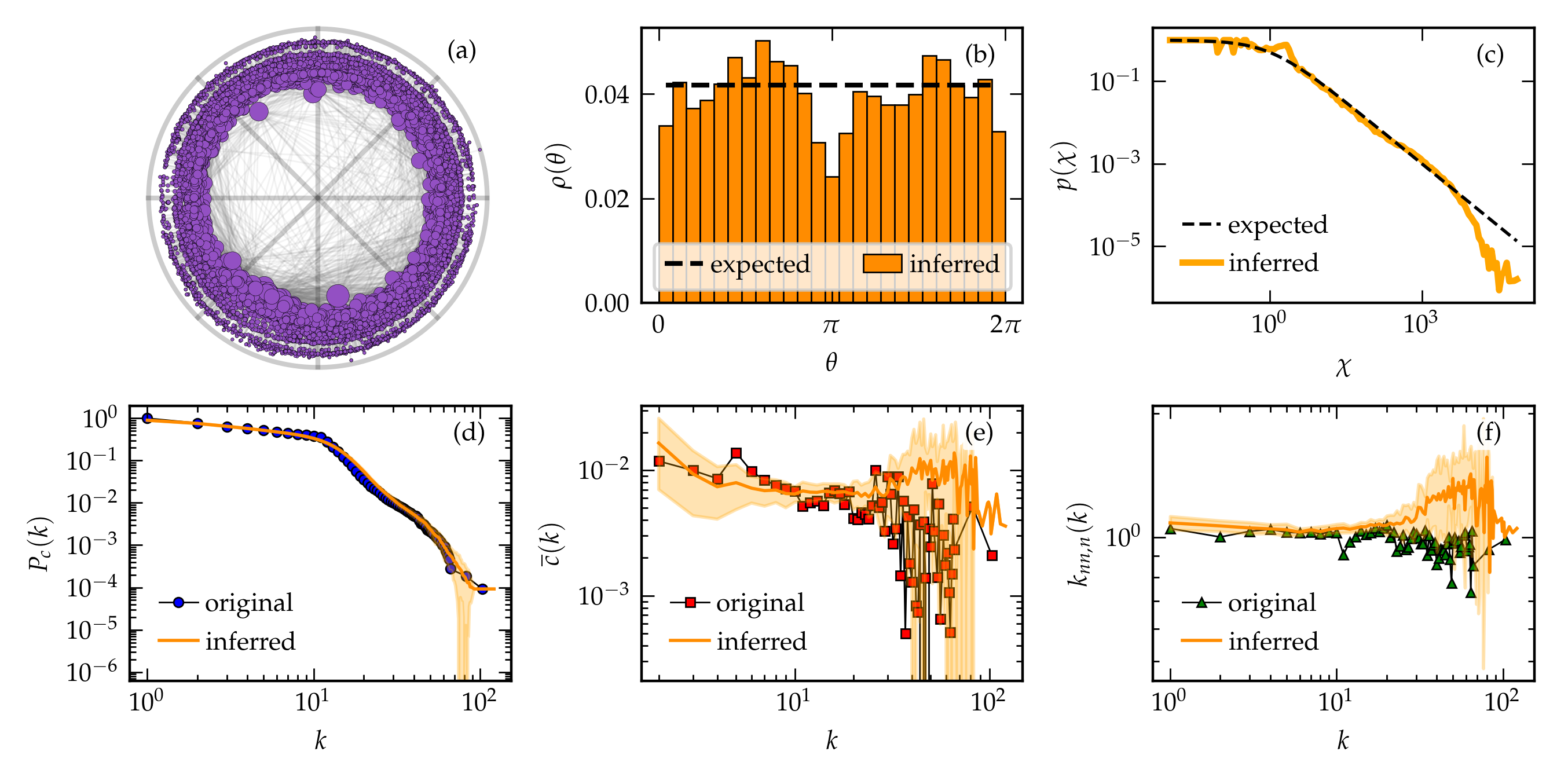}
	\vspace{-8mm}
	\caption{Summary of the results of Mercator for the Gnutella network. \textbf{(a)} Representation of the embedding in the hyperbolic plane as defined by the $\mathbb{H}^2$-model. The top 40\% most geometric edges are shown. \textbf{(b)} Comparison between the expected and inferred densities of nodes along the circle. \textbf{(c)} Comparison between the probability distribution as expected based on the model (expected) as well as the actual distribution based on the inferred coordinates (inferred). The reproduction of the topological properties is also given: \textbf{(d)} the complementary cumulative degree distribution, \textbf{(e)} the average local clustering coefficient per degree class and \textbf{(f)} the degree-degree correlations per degree class. The inferred results are obtained by generating 100 realizations of the $\mathbb{S}^1$-model based on the inferred coordinates. The orange shaded regions represent the $2\sigma$ confidence interval. }
	\label{Sfig:fooweb_baywet}
\end{figure}


\begin{figure}[b]
	\centering
	\includegraphics[width=1\textwidth]{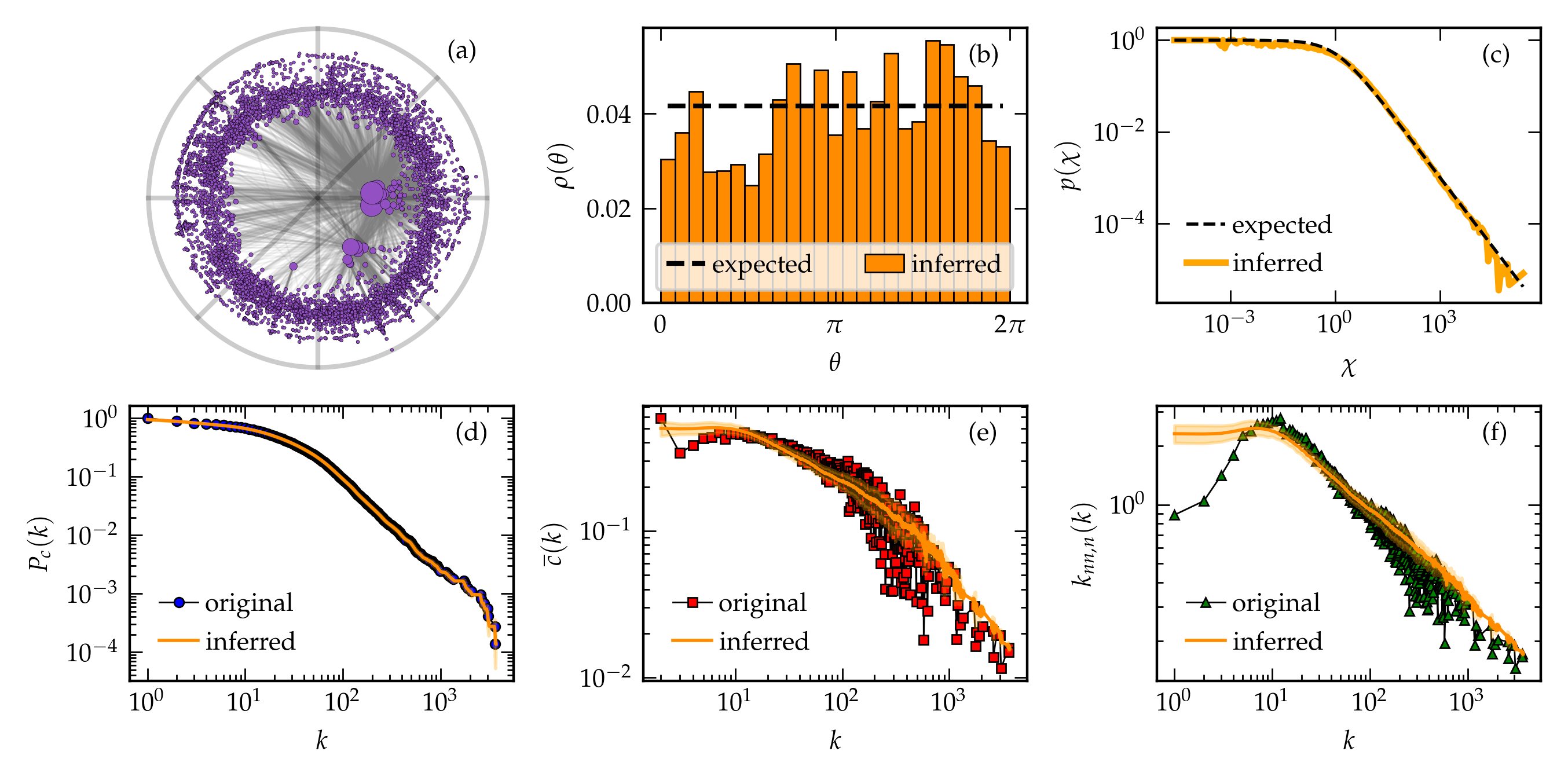}
	\vspace{-8mm}
	\caption{Summary of the results of Mercator for the PPI–S.cerevisiae network. \textbf{(a)} Representation of the embedding in the hyperbolic plane as defined by the $\mathbb{H}^2$-model. The top 3\% most geometric edges are shown. \textbf{(b)} Comparison between the expected and inferred densities of nodes along the circle. \textbf{(c)} Comparison between the probability distribution as expected based on the model (expected) as well as the actual distribution based on the inferred coordinates (inferred). The reproduction of the topological properties is also given: \textbf{(d)} the complementary cumulative degree distribution, \textbf{(e)} the average local clustering coefficient per degree class and \textbf{(f)} the degree-degree correlations per degree class. The inferred results are obtained by generating 100 realizations of the $\mathbb{S}^1$-model based on the inferred coordinates. The orange shaded regions represent the $2\sigma$ confidence interval. }
	\label{Sfig:fooweb_baywet}
\end{figure}


\begin{figure}[b]
	\centering
	\includegraphics[width=1\textwidth]{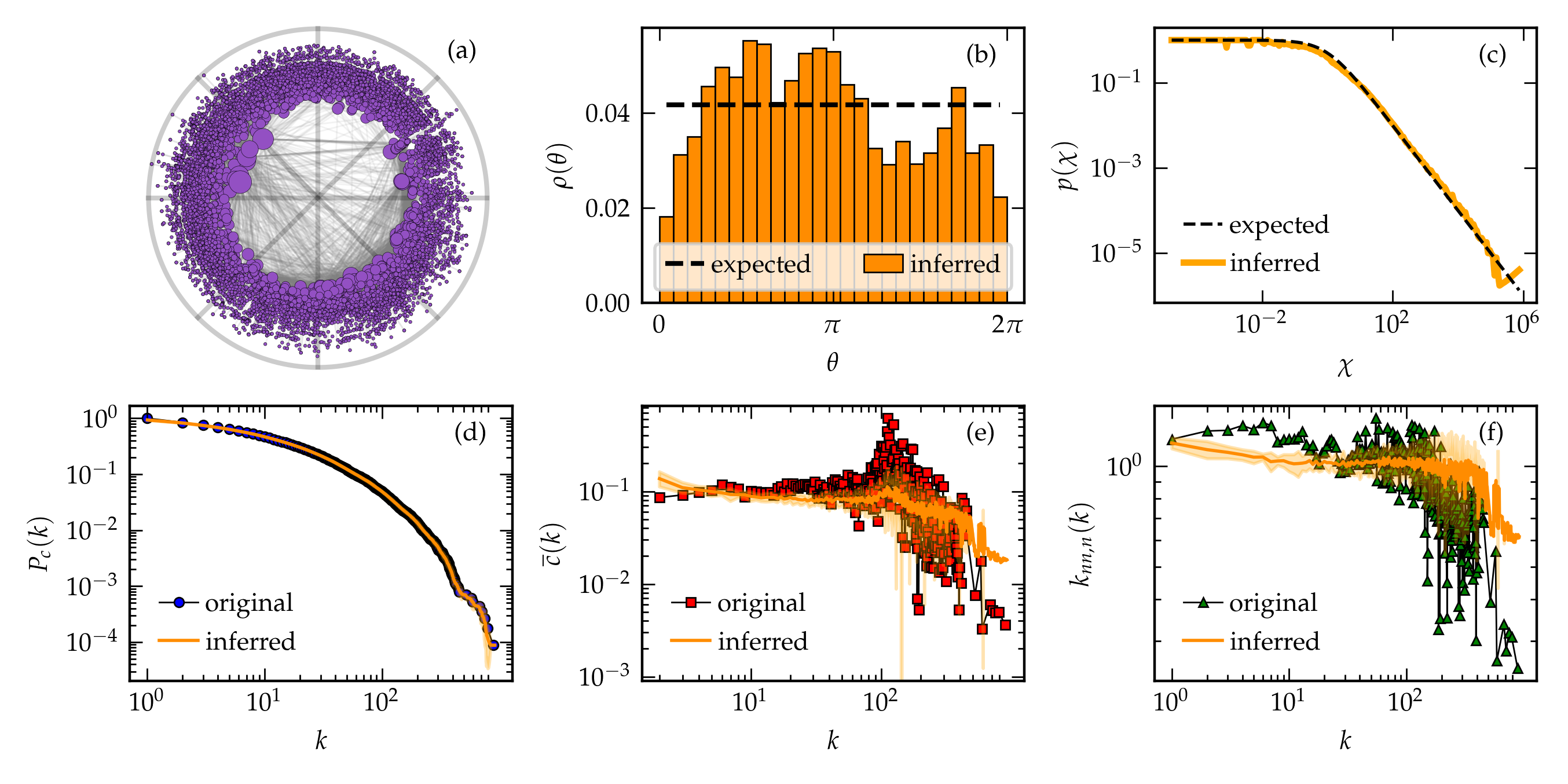}
	\vspace{-8mm}
	\caption{Summary of the results of Mercator for the PPI–D.melanogaster network. \textbf{(a)} Representation of the embedding in the hyperbolic plane as defined by the $\mathbb{H}^2$-model. The top 20\% most geometric edges are shown. \textbf{(b)} Comparison between the expected and inferred densities of nodes along the circle. \textbf{(c)} Comparison between the probability distribution as expected based on the model (expected) as well as the actual distribution based on the inferred coordinates (inferred). The reproduction of the topological properties is also given: \textbf{(d)} the complementary cumulative degree distribution, \textbf{(e)} the average local clustering coefficient per degree class and \textbf{(f)} the degree-degree correlations per degree class. The inferred results are obtained by generating 100 realizations of the $\mathbb{S}^1$-model based on the inferred coordinates. The orange shaded regions represent the $2\sigma$ confidence interval. }
	\label{Sfig:fooweb_baywet}
\end{figure}


\begin{figure}[b]
	\centering
	\includegraphics[width=1\textwidth]{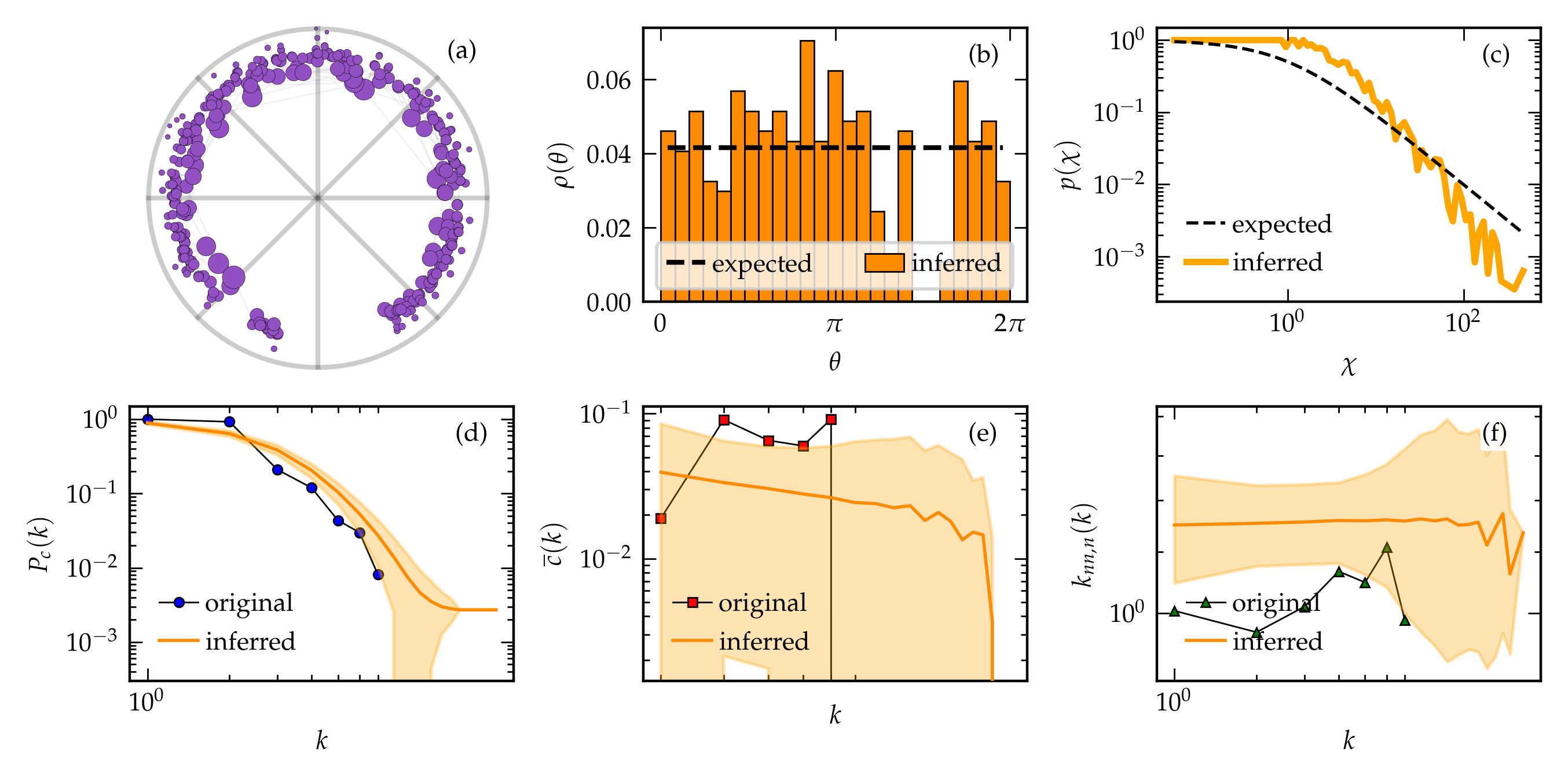}
	\vspace{-8mm}
	\caption{Summary of the results of Mercator for the Transport–London network. \textbf{(a)} Representation of the embedding in the hyperbolic plane as defined by the $\mathbb{H}^2$-model. The top 100\% most geometric edges are shown. \textbf{(b)} Comparison between the expected and inferred densities of nodes along the circle. \textbf{(c)} Comparison between the probability distribution as expected based on the model (expected) as well as the actual distribution based on the inferred coordinates (inferred). The reproduction of the topological properties is also given: \textbf{(d)} the complementary cumulative degree distribution, \textbf{(e)} the average local clustering coefficient per degree class and \textbf{(f)} the degree-degree correlations per degree class. The inferred results are obtained by generating 100 realizations of the $\mathbb{S}^1$-model based on the inferred coordinates. The orange shaded regions represent the $2\sigma$ confidence interval. }
	\label{Sfig:fooweb_baywet}
\end{figure}


\begin{figure}[b]
	\centering
	\includegraphics[width=1\textwidth]{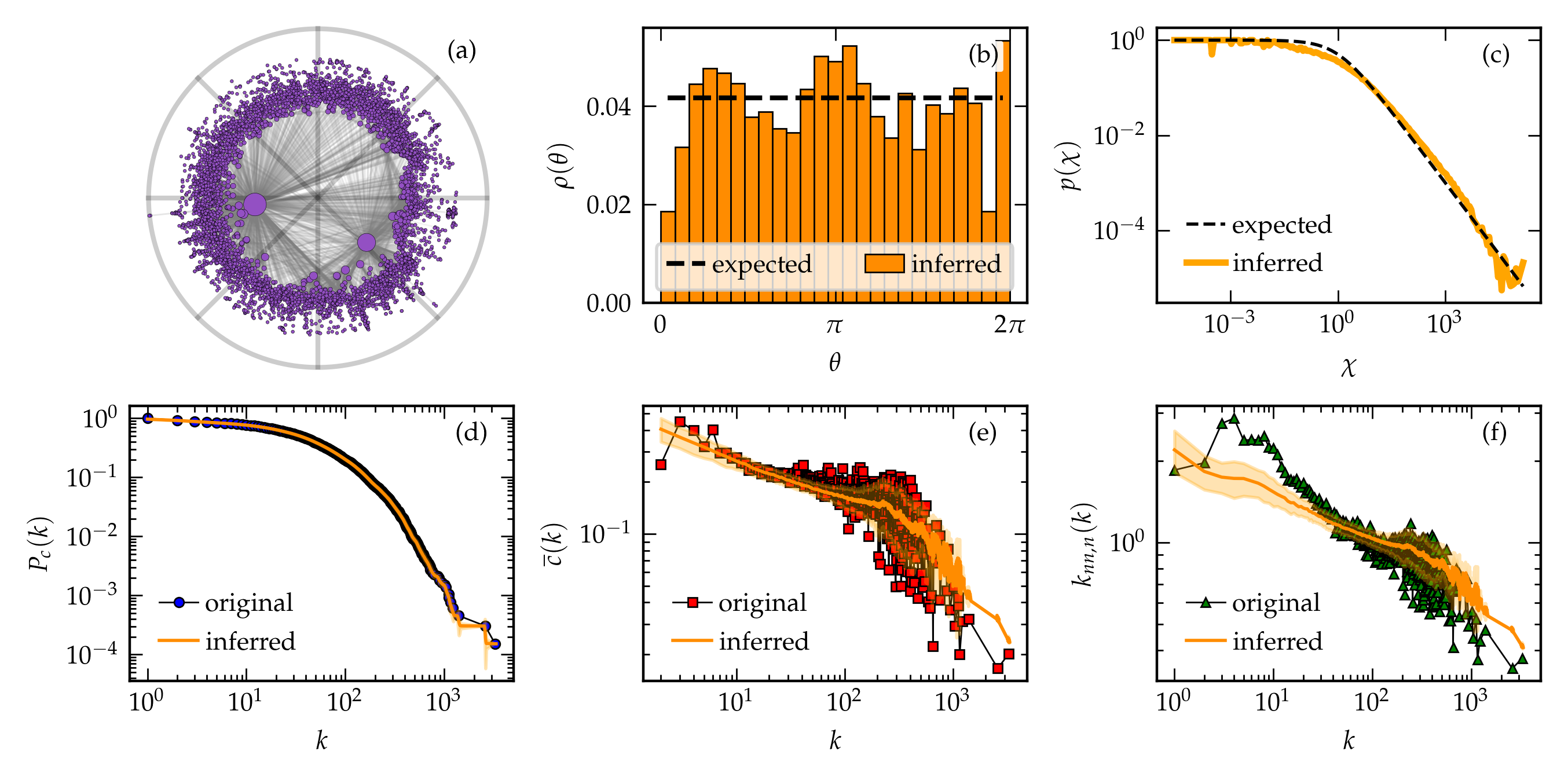}
	\vspace{-8mm}
	\caption{Summary of the results of Mercator for the GMP–S.cerevisiae network. \textbf{(a)} Representation of the embedding in the hyperbolic plane as defined by the $\mathbb{H}^2$-model. The top 5\% most geometric edges are shown. \textbf{(b)} Comparison between the expected and inferred densities of nodes along the circle. \textbf{(c)} Comparison between the probability distribution as expected based on the model (expected) as well as the actual distribution based on the inferred coordinates (inferred). The reproduction of the topological properties is also given: \textbf{(d)} the complementary cumulative degree distribution, \textbf{(e)} the average local clustering coefficient per degree class and \textbf{(f)} the degree-degree correlations per degree class. The inferred results are obtained by generating 100 realizations of the $\mathbb{S}^1$-model based on the inferred coordinates. The orange shaded regions represent the $2\sigma$ confidence interval. }
	\label{Sfig:fooweb_baywet}
\end{figure}


\begin{figure}[b]
	\centering
	\includegraphics[width=1\textwidth]{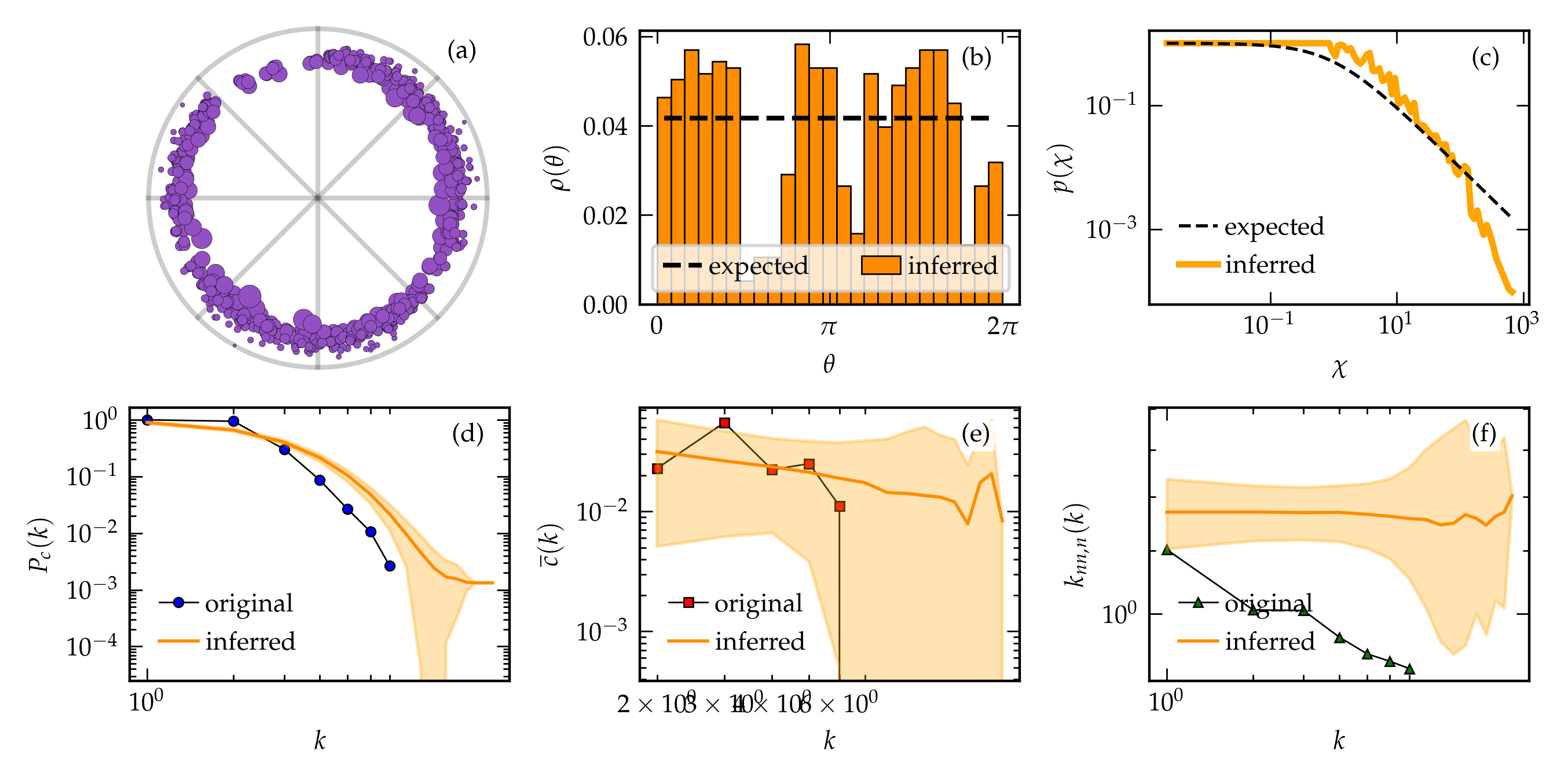}
	\vspace{-8mm}
	\caption{Summary of the results of Mercator for the Internet-PoP network. \textbf{(a)} Representation of the embedding in the hyperbolic plane as defined by the $\mathbb{H}^2$-model. The top 100\% most geometric edges are shown. \textbf{(b)} Comparison between the expected and inferred densities of nodes along the circle. \textbf{(c)} Comparison between the probability distribution as expected based on the model (expected) as well as the actual distribution based on the inferred coordinates (inferred). The reproduction of the topological properties is also given: \textbf{(d)} the complementary cumulative degree distribution, \textbf{(e)} the average local clustering coefficient per degree class and \textbf{(f)} the degree-degree correlations per degree class. The inferred results are obtained by generating 100 realizations of the $\mathbb{S}^1$-model based on the inferred coordinates. The orange shaded regions represent the $2\sigma$ confidence interval. }
	\label{Sfig:fooweb_baywet}
\end{figure}


\begin{figure}[b]
	\centering
	\includegraphics[width=1\textwidth]{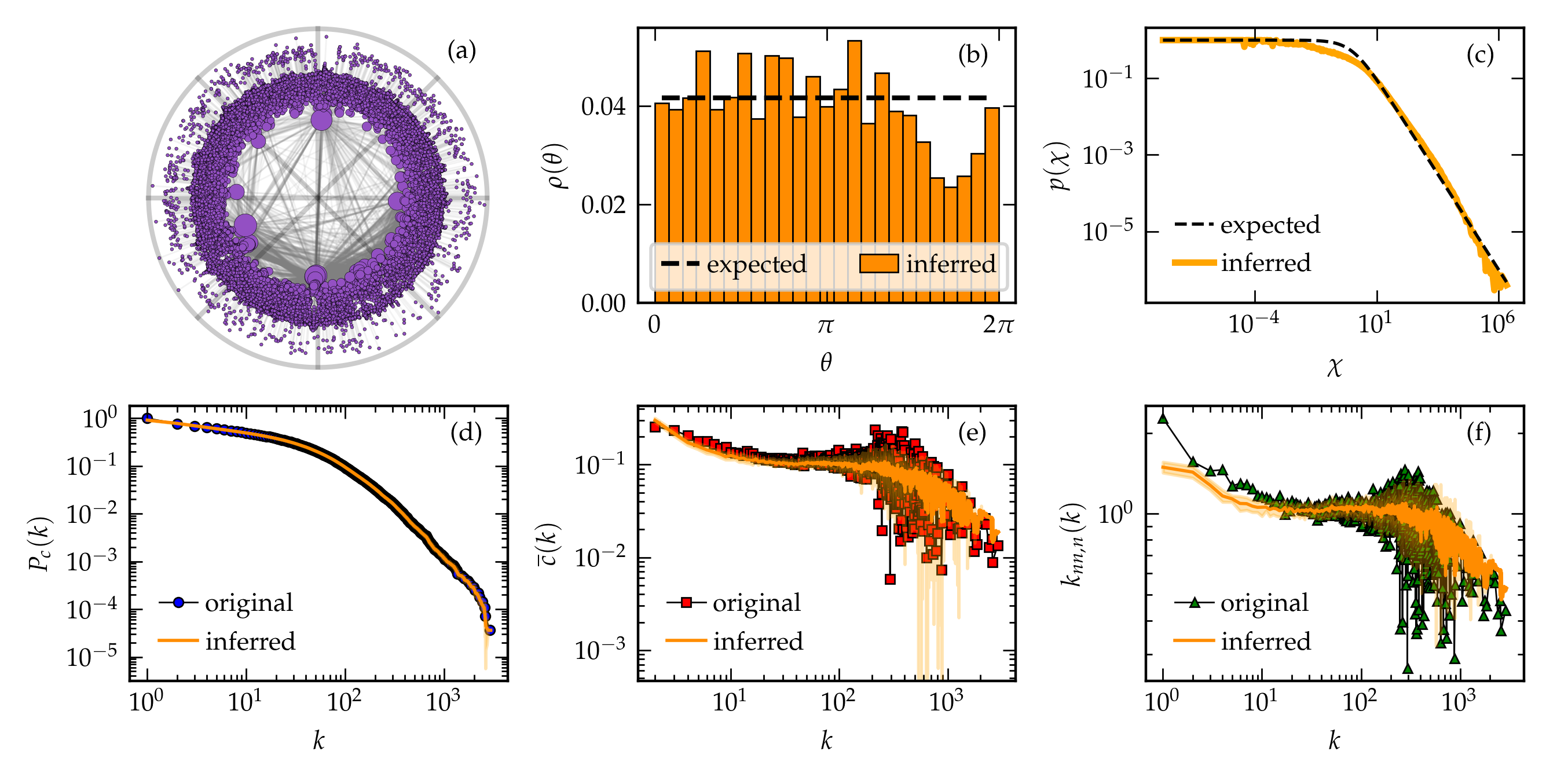}
	\vspace{-8mm}
	\caption{Summary of the results of Mercator for the PPI–H.sapiens network. \textbf{(a)} Representation of the embedding in the hyperbolic plane as defined by the $\mathbb{H}^2$-model. The top 5\% most geometric edges are shown. \textbf{(b)} Comparison between the expected and inferred densities of nodes along the circle. \textbf{(c)} Comparison between the probability distribution as expected based on the model (expected) as well as the actual distribution based on the inferred coordinates (inferred). The reproduction of the topological properties is also given: \textbf{(d)} the complementary cumulative degree distribution, \textbf{(e)} the average local clustering coefficient per degree class and \textbf{(f)} the degree-degree correlations per degree class. The inferred results are obtained by generating 100 realizations of the $\mathbb{S}^1$-model based on the inferred coordinates. The orange shaded regions represent the $2\sigma$ confidence interval. }
	\label{Sfig:fooweb_baywet}
\end{figure}


\begin{figure}[b]
	\centering
	\includegraphics[width=1\textwidth]{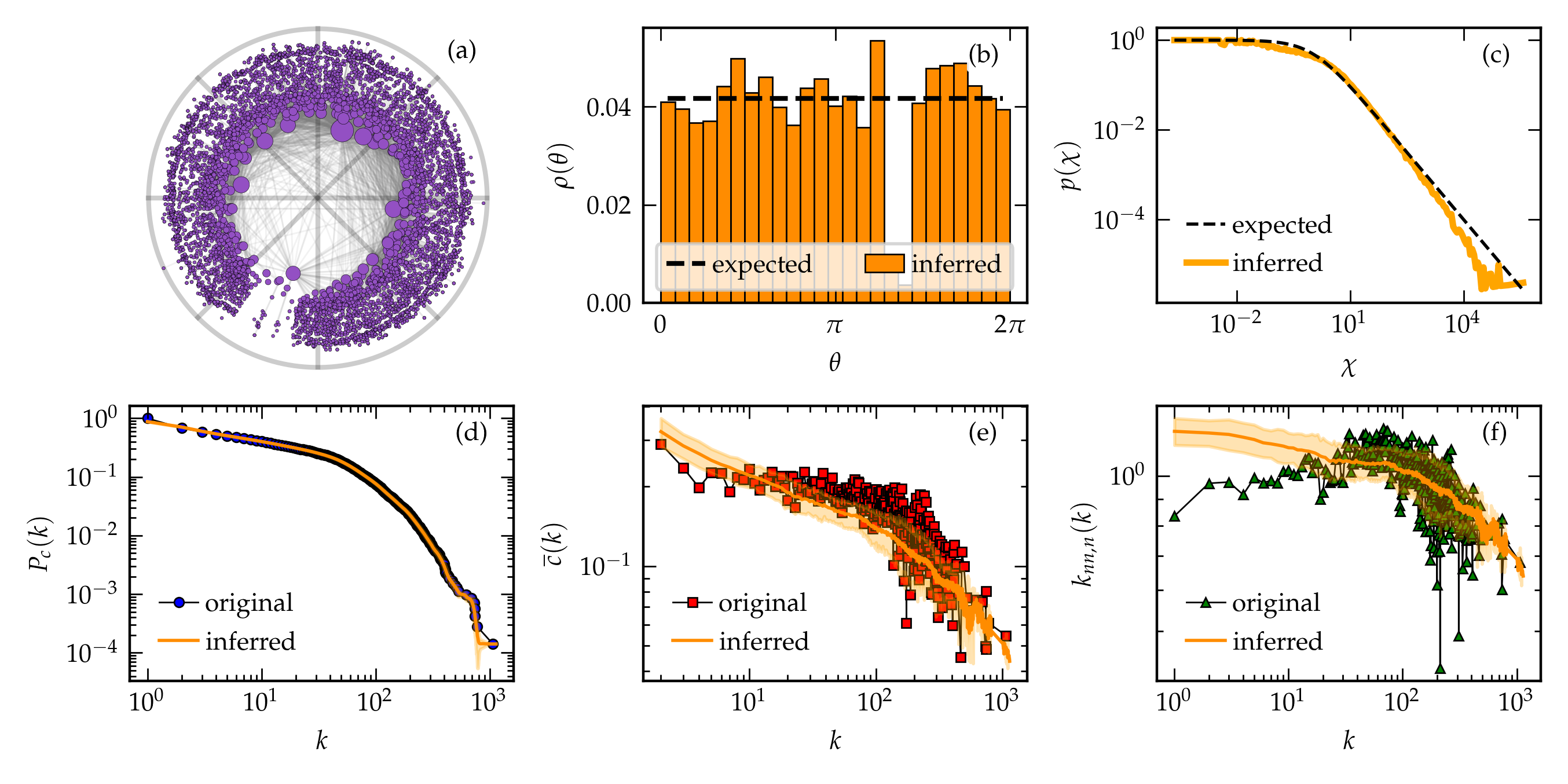}
	\vspace{-8mm}
	\caption{Summary of the results of Mercator for the WikiVote network. \textbf{(a)} Representation of the embedding in the hyperbolic plane as defined by the $\mathbb{H}^2$-model. The top 10\% most geometric edges are shown. \textbf{(b)} Comparison between the expected and inferred densities of nodes along the circle. \textbf{(c)} Comparison between the probability distribution as expected based on the model (expected) as well as the actual distribution based on the inferred coordinates (inferred). The reproduction of the topological properties is also given: \textbf{(d)} the complementary cumulative degree distribution, \textbf{(e)} the average local clustering coefficient per degree class and \textbf{(f)} the degree-degree correlations per degree class. The inferred results are obtained by generating 100 realizations of the $\mathbb{S}^1$-model based on the inferred coordinates. The orange shaded regions represent the $2\sigma$ confidence interval. }
	\label{Sfig:fooweb_baywet}
\end{figure}


\begin{figure}[b]
	\centering
	\includegraphics[width=1\textwidth]{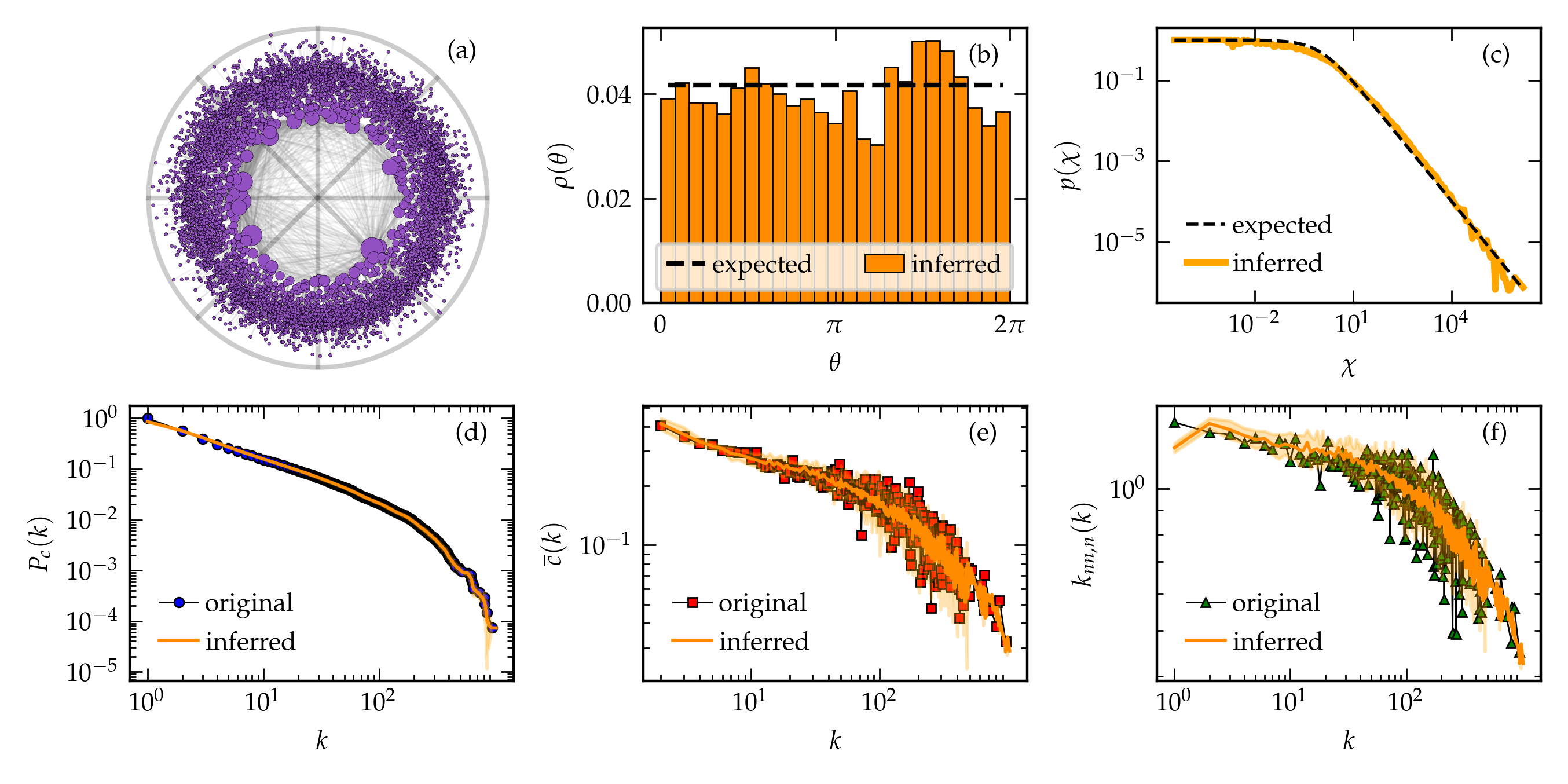}
	\vspace{-8mm}
	\caption{Summary of the results of Mercator for the MathOverflow network. \textbf{(a)} Representation of the embedding in the hyperbolic plane as defined by the $\mathbb{H}^2$-model. The top 5\% most geometric edges are shown. \textbf{(b)} Comparison between the expected and inferred densities of nodes along the circle. \textbf{(c)} Comparison between the probability distribution as expected based on the model (expected) as well as the actual distribution based on the inferred coordinates (inferred). The reproduction of the topological properties is also given: \textbf{(d)} the complementary cumulative degree distribution, \textbf{(e)} the average local clustering coefficient per degree class and \textbf{(f)} the degree-degree correlations per degree class. The inferred results are obtained by generating 100 realizations of the $\mathbb{S}^1$-model based on the inferred coordinates. The orange shaded regions represent the $2\sigma$ confidence interval. }
	\label{Sfig:fooweb_baywet}
\end{figure}